\documentclass[twocolumn,aps,prb,showpacs]{revtex4-1}
\usepackage{amsmath,amssymb,amsthm}
\usepackage{mathtools}
\usepackage{xspace}
\usepackage[usenames,dvipsnames]{xcolor}
\usepackage[colorlinks,hyperindex,breaklinks]{hyperref}
\usepackage{braket}
\usepackage{hyphenat}
\usepackage{bbold}
\usepackage{enumerate}
\usepackage{tabu,booktabs}
\usepackage{tabularx}
\usepackage[verbose]{placeins}
\usepackage[hang,flushmargin]{footmisc}
\usepackage{comment}

\usepackage[british]{babel}

\usepackage{color}
\def\tr{\textcolor{red}}
\def\tr{}
\usepackage{lipsum}
\def\ptfecocu{Pt\,$\vert$\,Fe$_x$Co$_{1-x}$\,$\vert$\,Cu~}
\newcommand{\sw}[1]{\textbf{\textcolor{blue}{#1}}}
\renewcommand{\sw}{}

\renewcommand{\arraystretch}{1.6}

\renewcommand{\vec}[1]{\mathbf{#1}}

\newcommand{\ea}{\textit{et al.}}

\begin{document}

\title{A fully relativistic description of spin-orbit torques
by means of linear response theory}

\author{S. Wimmer}
\email{sebastian.wimmer@cup.uni-muenchen.de}
\author{K. Chadova}
\author{M. Seemann}
\author{D. K\"odderitzsch}
\author{H. Ebert}
\email{hubert.ebert@cup.uni-muenchen.de}

\affiliation{Department Chemie/Phys.\ Chemie, 
            Ludwig-Maximilians-Universit\"at
           M\"unchen, Germany}

\date{\today}


\begin{abstract}
Symmetry and magnitude of spin-orbit torques (SOT), i.e., current-induced
torques on the magnetization of systems lacking inversion symmetry, are
investigated in a fully relativistic linear response framework based on the
Kubo formalism.  By applying all space-time symmetry operations contained in
the magnetic point group of a solid to the relevant response coefficient, the
torkance expressed as torque-current correlation function, restrictions to the
shape of the direct and inverse response tensors are obtained. These are shown
to apply to the corresponding thermal analogues as well, namely the direct and
inverse thermal SOT in response to a temperature gradient or heat current.
Using an implementation of the Kubo-Bastin formula for the torkance into a
first-principles multiple-scattering Green's function framework and accounting
for disorder effects via the so-called coherent potential approximation (CPA),
all contributions to the SOT in pure systems, dilute as well as concentrated
alloys can be treated on equal footing. This way, material specific values for
all torkance tensor elements in the fcc (111) trilayer alloy system
\ptfecocu are obtained over a wide concentration
range and discussed in comparison to results for electrical and spin
conductivity, as well as to previous work -- in particular concerning symmetry
w.r.t.\ magnetization reversal and the nature of the various contributions. 
\end{abstract}


\maketitle

\renewcommand{\arraystretch}{1.6}
\newcommand{\dd}{\mathrm d}
\newcommand{\del}{\partial}

\newcommand{\matr}[1]{\mathbf{\underline{#1}}}
\newcommand{\matrg}[1]{\underline{\boldsymbol{#1}}}

\newcommand{\ee}{\text e}
\newcommand{\ii}{\text i}

\renewcommand{\vec}[1]{\mathbf{#1}}
\newcommand{\spacegroup}[1]{$#1$}

\newcommand{\tableheading}[1]{\textbf{#1}}

\newcommand{\tensordir}{../sym/tmp/}
\newcommand{\matrixtablelinesep}{0.55mm}
\newcommand{\matrixtablearraystretch}{0.9}
\newcommand{\matrixtablearraycolsep}{0.3pt}

\newcommand{\TR}{\mbox{Tr}}
\newcommand{\SIGTEN}{\mbox{$\matrg \sigma$}}
\newcommand{\TORTEN}{\mbox{$\matrg t$}}

\section{Introduction \label{sec:I}}

Spin-orbit torques (SOT), denoting the response of a magnetization to an
electric current by changing its orientation, have evolved from a theoretical
conjecture\cite{MZ08,MZ09,GM09a} via experimental
verification\cite{COL+09,MGA+10,PWB+10} to their imminent technological
application in SOT-MRAM devices\cite{GMG+10} in a remarkably short period of
time. This can be attributed to the fact that unlike most\footnote{\sw{For
alternatives to the SOT for \emph{control of magnetism by electric fields} see
for example the same-titled review by Matsukara \ea\cite{MTO15}}} other
ways of defined manipulation of magnetic moments it does not require external
magnetic fields or auxiliary magnetic layers, offering an enormous advantage
concerning information storage density, non-volatility and
scalability.\cite{MGG+11} The combined effect of spin-orbit interaction and
exchange coupling in systems lacking inversion symmetry offers thus the
possibility to switch the magnetization of spintronics devices by applying an
electric current. Unlike its close relative, the spin-transfer torque
mechanism,\cite{Slo96,Ber96} it does not rely on the presence of a
``polarizer'' magnetic layer, allowing for much simpler device architecture and
reducing necessary critical current densities.\cite{GM11} The intrinsic
relativistic spin-orbit interaction can transfer orbital to spin angular
momentum in a magnetic material having a suitable structure, leading to an
effective magnetic field exerting a torque on the magnetization.

%
Recent experiments\cite{GMA+13,KSH+13,QDN+14} where able to measure the SOT
directly as a function of magnetization direction, whereas earlier evidence has
only been indirect.\cite{COL+09,MGA+10,PWB+10,MGG+11,MMS+11,LLG+12,LPL+12} Two
symmetrically distinct contributions to the SOT could be observed this way, one
being even and the other odd with respect to magnetization reversal.
To lowest order in the magnetization direction $\hat{\bf m}$, the even torque
in response to an in-plane current $\vec{j}$ was found to go by $\hat{\bf m}
\times (\vec{j} \times \hat{\bf m})$ while the odd one scales with  $\hat{\bf
m} \times \vec{j}$.
While initial work on spin-orbit torques was focused on transition metal
FM\,$\vert$\,NM bilayers\cite{MGA+10,PWB+10} or dilute magnetic
semiconductors,\cite{COL+09} recent
experiments\cite{RKH+15,TCM+15,ZJF+15,FZD+16} demonstrate the possibility to
switch ferromagnetic moments by SOTs originating from antiferromagnets, as
predicted theoretically.\cite{ZGV+14} On the ferromagnetic side it could be
shown that magnetic insulators can be switched by SOTs as well.\cite{HAH+14}
Exploiting the large spin-orbit coupling of Bismuth and the pronounced
ferromagnetism of Cr-doped Bi$_x$Sb$_{1-x}$Te$_3$, topological insulator
heterostructures were shown to be promising candidates for SOT-based memory and
logic devices.\cite{FUK+14,WDB+15}
%
Although the nature of the spin-orbit torque certainly is not yet fully
understood in all details, its ability to deterministically switch magnetic
moments, without the need for external magnetic fields,\cite{YUF+14} has been
demonstrated beyond doubt.  Currently, experimental research is already heading
towards fully functional devices,\cite{CBD+14,PJP+15} en route exposing further
interesting aspects of SOTs.\cite{SJL+16,YCJ+16}

%
%
Early theoretical work on the SOT in bilayer systems proposed two distinct
mechanisms, namely a \if{}field-like\fi torque arising from the Rashba effect
at the asymmetric interface,\cite{MZ09,MR09,KMLL12,TNL13} and a spin transfer
torque due to the spin current generated in the heavy-metal layer by the spin
Hall effect\if{}, also termed damping-like due to its
symmetry\fi.\cite{LMRB11,LLG+12,LPL+12}
%
%
%
The ``Rashba''-torque was initially found to be dominated by a field-like
component\cite{GM09,MR09,HLL+13} being odd w.r.t.\ magnetization reversal,
whereas the ``spin~Hall''-torque was believed to consist mostly of an even
(anti-)damping- or spin transfer-like
contribution.\cite{MGG+11,LLG+12,LPL+12,HLL+13}  This picture has however
turned out to be too simple,\cite{GMA+13,KSH+13} as these mechanisms appear to
be only the limiting cases of a more complex scenario,\cite{Man12,GMA+13}
involving terms of higher-order in the magnetization direction\cite{GMA+13} and
in addition an intrinsic contribution, arising from the band structure in a
single ferromagnetic layer alone.\cite{KSF+14,KLLS15}
%
%
First-principles calculations of the torkance tensor\cite{HLL+13a,FBM14,FBM14a} can be
used to help to disentangle the various contributions by providing
model-independent material parameters. The pioneering works of Freimuth
\ea\cite{FBM14,FBM14a} demonstrated this for FM\,$\vert$\,NM~bilayer
systems using the Kubo linear response formalism to calculate layer-resolved
torkances.

It has been noted quite early on,\cite{HBT10,LMRB11,KMLL12,TNL13} that there
exists of course an Onsager reciprocal to the spin-orbit torque, i.e., by
interchanging perturbation (electric field or charge current) and response
(torque on the magnetization) one arrives at the inverse spin-orbit torque
(ISOT), describing the electric field induced by magnetization
dynamics.\cite{CHI+15} The reciprocity of the two, SOT and ISOT, has been
discussed recently in great detail by Freimuth \ea,\cite{FBM15} who noted that
both phenomena can be described by the torkance tensor. In this work, by
performing a symmetry analysis of the linear response expressions describing
SOT and ISOT, we will give explicit tensor shapes for both properties in terms
of the torkance, thereby demonstrating, where applicable, the presence and
exact form of their reciprocity. As will be demonstrated, these shapes remain
unchanged when replacing the electric field by a temperature gradient, giving a
justification for the use of a Mott-like expression for direct and inverse
thermal SOT discussed by G\'eranton \ea \cite{GFBM15} and Freimuth
\ea\cite{FBM16}.

The present paper focuses on two aspects of the spin-orbit torque that have, to
our knowledge, not been studied before. Firstly, an extensive symmetry analysis
based on group-theoretical grounds and not restricted to special cases is
preformed that allows determining the tensor shapes of both direct and inverse
SOT from their respective Kubo linear response expressions, based on the
magnetic point group alone. Secondly, by making use of the Coherent Potential
Approximation (CPA) within multiple scattering theory the possibility to study
the concentration-dependence of the torkance in alloys is demonstrated, thereby
opening the route for a materials design approach to the SOT.

This paper is organized as follows: In Section~\ref{sec:F} we introduce the
underlying linear response formalism used to calculate the torkance tensor,
discuss its implementation into a multiple scattering framework, with
particular emphasis on the treatment of disorder, and finally outline the
application of symmetry considerations leading to restrictions to the tensor
shapes of both, direct and inverse spin-orbit torques. The outcome of this
group-theoretical analysis for all magnetic point groups allowing for the
existence of a finite magnetization will be presented together with
corresponding results for the electrical and spin conductivity tensors. In
Section~\ref{sec:R} we present the results of our numerical investigations on a
\ptfecocu trilayer system, highlighting the impact of disorder
effects (impurity scattering) on the various contributions to the torkance. By
comparing concentration-dependent results for the torkance tensor with such for
the spin Hall conductivity we will discuss their (partial) interconnection.
Finally, contact will be made to previous work, in particular concerning the
separation of the torkance into contributions based on the structure of the
linear response expression (Fermi sea and Fermi surface terms) \if, based on
scattering theory (on-site vs.\ backscattering contributions)\fi and on symmetry
arguments (even or odd symmetry w.r.t.\ magnetization reversal). We conclude with
a summary of the presented and an outlook on future work in Section~\ref{sec:C}.

\section{Formalism \label{sec:F}}

A well-known application of  Kubo's linear response formalism is the derivation
of an expression for the electrical conductivity tensor \SIGTEN\ that
describes the electric current density $\vec{j}=\SIGTEN \, \vec{E}$ in
response to an electric field $\vec{E}$.  In analogy one can derive an
expression for the torkance tensor \TORTEN\ that gives the torque
$\vec{T}=\TORTEN \, \vec{E}$ as a response to $\vec{E}$.\cite{FBM14,FBM14a}
Replacing the operator $\hat{j}_{\mu}$ representing  the component $\mu$ of the
current density by the operator  $\hat{T}_{\mu}$ for the torque one can
straightforwardly adopt the derivation of the so-called Kubo-Bastin formula for
\SIGTEN\ ,\cite{BLBN71,CB01a} leading to a corresponding expression  for the
torkance \TORTEN:\cite{FBM14remark}
%
\begin{widetext}
\begin{eqnarray}
  \label{eq:torkance-Bastin1}
  \nonumber
  t_{\mu\nu}
  &=&
    - \frac{\hbar }{4\pi V}\int_{-\infty}^{\infty}d\varepsilon
    \frac{df(\varepsilon)}{d\varepsilon}
    \TR
    \left<
      \hat{T}_{\mu}(G^{+}-G^{-})  \hat{j}_{\nu} G^{-}
      -\hat{T}_{\mu} G^{+} \hat{j}_{\nu}(G^{+}-G^{-})
    \right>
  \\
  &&
    + \frac{\hbar }{4\pi V}\int_{-\infty}^{\infty}d\varepsilon
    f(\varepsilon)
    \TR
    \left<
        \hat{T}_{\mu}G^{+}\hat{j}_{\nu}
        \frac{dG^{+}}{d\varepsilon}
        -\hat{T}_{\mu}\frac{dG^{+}}{d\varepsilon}  \hat{j}_{\nu} G^{+}
      -
        \mbox{``}\left(
        G^{+} \rightarrow G^{-}
        \right)\mbox{``}
    \right>
    \label{eq:torkance-Bastin2}
\; ,
\end{eqnarray}
\end{widetext}
%
\sw{where $V$ is the volume of the unit cell and} $f(E)$ is the Fermi
distribution function. This implies that in the limit $T \rightarrow 0$~K for
the temperature the first term in Eq.~\eqref{eq:torkance-Bastin1} has to be
evaluated only for the Fermi energy $E_{\rm F}$ (Fermi surface term $
t_{\mu\nu}^{I}$), while the second one requires an integration over the
occupied part of the valence band  (Fermi sea term $ t_{\mu\nu}^{II}$).

The operator $\hat{j}_{\nu}= - |e| c \alpha_\nu$ in
Eq.~\eqref{eq:torkance-Bastin1} represents the perturbation due to the electric
field component $E_\nu$.  Adopting a fully relativistic formulation to account
coherently for the impact of SOC,  $\hat{j}_{\nu}$ is expressed by the
corresponding velocity operator  $\hat{v}_{\nu}= c \alpha_\nu$, where $c$ is
the speed of light and  $\alpha_\nu$ is one  of the standard $4 \times 4 $
Dirac matrices.\cite{Ros61} The torque operator $\hat{T}_\mu$ on the other hand
represents the change of the magnetization component  $m_\mu$ with time in
response to the electric field $\vec{E}$.  Accordingly,  $\hat{T}_\mu$  may be
expressed by the partial derivative of the Dirac Hamiltonian $\cal H $ with
respect to the component $ u_\mu$ of the normalized magnetisation  $\vec m /
|\vec m|$:\cite{EMKK11}
%
\begin{eqnarray}
\hat{T}_{\mu}
& = & 
\frac{\partial}{\partial u_\mu}\hat{\cal H}
 \nonumber \\
& = & \beta
\sigma_{\mu}B_{xc}(\vec{r})
\label{matrix-element} \;.
\end{eqnarray}
%
For  the second line use has been made of the specific form of  $\hat{\cal H}$
for a magnetic solid within the framework of local spin density formalism
(LSDA) where $B_{xc}(\vec{r})$ stands for the difference in the exchange
potential for electrons with spin up and down \cite{MV79} and $\sigma_{\mu}$ is
one of the $4 \times 4 $ Pauli spin matrices.\cite{Ros61}

In  Eq.~\eqref{eq:torkance-Bastin1} the electronic structure is represented in
terms of the retarded and advanced Green functions $G^+(E)$ and $G^-(E)$,
respectively.  Using this approach has the big advantage that one can deal
straightforwardly with disordered systems.  Considering for example chemical
disorder the brackets $\langle ... \rangle$ in  Eq.~\eqref{eq:torkance-Bastin1}
stand for the configurational average \sw{in} a disordered alloy. For the
applications presented below relativistic multiple scattering theory was used
to determine the Green function.\cite{Ebe00,EKM11} The \sw{average over} alloy
configurations was determined by means of the Coherent Potential Approximation
(CPA) alloy theory as done in the context of the electrical
conductivity,\cite{But85,LKE10b} spin conductivity \cite{LGK+11} and Gilbert
damping parameter.\cite{EMKK11}  This implies in particular that the so-called
vertex corrections, that ensure that the proper average $\langle \hat{T}_\mu
G^\pm \hat{j}_\nu G^\pm \rangle $ is taken instead of the simpler one $\langle
\hat{T}_\mu G^\pm \rangle  \langle \hat{j}_\nu G^\pm \rangle$, are included in
the calculations.

\medskip

Expressing the electric field induced torque by means of linear response
formalism \sw{allows investigating} straightforwardly the condition for which the
SOT may show up or not.  This can be done using a scheme  worked out by Kleiner
\cite{Kle66} and extended recently by Seemann \ea.\cite{SKWE15} Making use of
the behavior of the torque operator $\hat T_\mu$ and of the current density
operator $j_{\nu}$ under symmetry operations one is led to the relations that
restrict the shape of the torkance tensor \TORTEN:
%
\begin{eqnarray}
\label{sym-unitary}
t_{\mu \nu }
 &= &
\sum_{\kappa \lambda} t_{\kappa \lambda}
 D(R)_{\kappa\mu } D(R)_{ \lambda\nu }\det(R)
\\
\label{sym-anti-unitary}
t_{\mu \nu }
& =& 
\sw{-}\sum_{\kappa \lambda} t'_{ \lambda\kappa}
 D(R)_{\kappa\mu }^* D(R)_{ \lambda\nu }^*\det(R) 
\; ,
\end{eqnarray}
%
where $ \matrg{D}(R) $ is the $3 \times 3 $ transformation matrix associated
with the pure spatial operation $R$ and $\det(R) $ is the corresponding
determinant of that matrix.  In Eq.~\eqref{sym-unitary} only unitary pure
spatial symmetry operations are considered, while in
Eq.~\eqref{sym-anti-unitary}  anti-unitary operations are considered that
involve apart from the spatial operation $R$ also the  time reversal operation.
As a consequence   Eq.~\eqref{sym-anti-unitary} relates the torkance  tensor
\TORTEN\ with the tensor $\matrg t'$ that is associated with the effect inverse
to the SOT, i.e.,  Eq.~\eqref{sym-anti-unitary} is \sw{equivalent to an}
Onsager relation for  \TORTEN.

Considering  Eq.~\eqref{sym-unitary} for all symmetry operations of a magnetic
point group, the corresponding symmetry-allowed shape of the direct and inverse
torkance tensors, \TORTEN\  and \TORTEN$^\prime$, can be determined.
Tables~\ref{TAB-TORTENb} to \ref{TAB-TORTENc2c} give the results for all
magnetic point groups leading to a non-vanishing  torkance  tensor.
\sw{
In addition the tensor shapes for electrical and spin conductivity for polarization
along the principal axis are given for the respective magnetic Laue groups
obtained by adding the spatial inversion operation.\cite{Kle66,SKWE15}
Naturally, this leads to redundancies since different magnetic point groups
have the same magnetic Laue group. 
}

\sw{
Magnetic symmetry groups that allow for a
finite magnetization in general can be subdivided into two
categories,\cite{Kle66,SKWE15} one without any time-reversal symmetry, neither
as an operation on its own nor in combination with a spatial operation,
category b) (Table~\ref{TAB-TORTENb}), the other containing time-reversal only
connected with a spatial operation, category c)
(Tables~\ref{TAB-TORTENc}--\ref{TAB-TORTENc2c}).
}
%
%
\begin{table*}
  \tabulinesep\matrixtablelinesep
  \renewcommand{\arraystretch}{\matrixtablearraystretch}
  \setlength\arraycolsep{\matrixtablearraycolsep}
  \centering
\begin{tabular}{ccccccc}
\tableheading{magnetic point group} & \tableheading{\TORTEN} & \tableheading{\TORTEN$^\prime$} & \tableheading{magnetic Laue group} & \tableheading{\SIGTEN} & \tableheading{\SIGTEN$^k$} \\[4mm]
\spacegroup{1}
&
\begin{math} \begin{pmatrix}
t_{xx} & t_{xy} & t_{xz} \\
t_{yx} & t_{yy} & t_{yz} \\
t_{zx} & t_{zy} & t_{zz}
\end{pmatrix} \end{math}
&
\begin{math} \begin{pmatrix}
t^\prime_{xx} & t^\prime_{xy} & t^\prime_{xz} \\
t^\prime_{yx} & t^\prime_{yy} & t^\prime_{yz} \\
t^\prime_{zx} & t^\prime_{zy} & t^\prime_{zz}
\end{pmatrix} \end{math}
&
\spacegroup{\bar 1}
&
\begin{math} \begin{pmatrix}
\sigma_{xx} & \sigma_{xy} & \sigma_{xz} \\
\sigma_{yx} & \sigma_{yy} & \sigma_{yz} \\
\sigma_{zx} & \sigma_{zy} & \sigma_{zz}
\end{pmatrix} \end{math}
&
\begin{math} \begin{pmatrix}
\sigma_{xx}^{z} & \sigma_{xy}^{z} & \sigma_{xz}^{z} \\
\sigma_{yx}^{z} & \sigma_{yy}^{z} & \sigma_{yz}^{z} \\
\sigma_{zx}^{z} & \sigma_{zy}^{z} & \sigma_{zz}^{z}
\end{pmatrix} \end{math}
\\[3ex]
\spacegroup{2}
&
\begin{math} \begin{pmatrix}
t_{xx} & 0 & t_{xz} \\
0 & t_{yy} & 0 \\
t_{zx} & 0 & t_{zz}
\end{pmatrix} \end{math}
&
\begin{math} \begin{pmatrix}
t^\prime_{xx} & 0 & t^\prime_{xz} \\
0 & t^\prime_{yy} & 0 \\
t^\prime_{zx} & 0 & t^\prime_{zz}
\end{pmatrix} \end{math}
&
\spacegroup{2/m}
&
\begin{math} \begin{pmatrix}
\sigma_{xx} & 0 & \sigma_{xz} \\
0 & \sigma_{yy} & 0 \\
\sigma_{zx} & 0 & \sigma_{zz}
\end{pmatrix} \end{math}
&
\begin{math} \begin{pmatrix}
\sigma_{xx}^{y} & 0 & \sigma_{xz}^{y} \\
0 & \sigma_{yy}^{y} & 0 \\
\sigma_{zx}^{y} & 0 & \sigma_{zz}^{y}
\end{pmatrix} \end{math}
\\[3ex]
\spacegroup{m}
&
\begin{math} \begin{pmatrix}
0 & t_{xy} & 0 \\
t_{yx} & 0 & t_{yz} \\
0 & t_{zy} & 0
\end{pmatrix} \end{math}
&
\begin{math} \begin{pmatrix}
0 & t^\prime_{xy} & 0 \\
t^\prime_{yx} & 0 & t^\prime_{yz} \\
0 & t^\prime_{zy} & 0
\end{pmatrix} \end{math}
&
\spacegroup{2/m}
&
\begin{math} \begin{pmatrix}
\sigma_{xx} & 0 & \sigma_{xz} \\
0 & \sigma_{yy} & 0 \\
\sigma_{zx} & 0 & \sigma_{zz}
\end{pmatrix} \end{math}
&
\begin{math} \begin{pmatrix}
\sigma_{xx}^{y} & 0 & \sigma_{xz}^{y} \\
0 & \sigma_{yy}^{y} & 0 \\
\sigma_{zx}^{y} & 0 & \sigma_{zz}^{y}
\end{pmatrix} \end{math}
\\[3ex]
\spacegroup{222}
&
\begin{math} \begin{pmatrix}
t_{xx} & 0 & 0 \\
0 & t_{yy} & 0 \\
0 & 0 & t_{zz}
\end{pmatrix} \end{math}
&
\begin{math} \begin{pmatrix}
t^\prime_{xx} & 0 & 0 \\
0 & t^\prime_{yy} & 0 \\
0 & 0 & t^\prime_{zz}
\end{pmatrix} \end{math}
&
\spacegroup{mmm}
&
\begin{math} \begin{pmatrix}
\sigma_{xx} & 0 & 0 \\
0 & \sigma_{yy} & 0 \\
0 & 0 & \sigma_{zz}
\end{pmatrix} \end{math}
&
\begin{math} \begin{pmatrix}
0 & \sigma_{xy}^{z} & 0 \\
\sigma_{yx}^{z} & 0 & 0 \\
0 & 0 & 0
\end{pmatrix} \end{math}
\\[3ex]
\spacegroup{mm2}
&
\begin{math} \begin{pmatrix}
0 & t_{xy} & 0 \\
t_{yx} & 0 & 0 \\
0 & 0 & 0
\end{pmatrix} \end{math}
&
\begin{math} \begin{pmatrix}
0 & t^\prime_{xy} & 0 \\
t^\prime_{yx} & 0 & 0 \\
0 & 0 & 0
\end{pmatrix} \end{math}
&
\spacegroup{mmm}
&
\begin{math} \begin{pmatrix}
\sigma_{xx} & 0 & 0 \\
0 & \sigma_{yy} & 0 \\
0 & 0 & \sigma_{zz}
\end{pmatrix} \end{math}
&
\begin{math} \begin{pmatrix}
0 & \sigma_{xy}^{z} & 0 \\
\sigma_{yx}^{z} & 0 & 0 \\
0 & 0 & 0
\end{pmatrix} \end{math}
\\[3ex]
\spacegroup{4}
&
\begin{math} \begin{pmatrix}
t_{xx} & t_{xy} & 0 \\
-t_{xy} & t_{xx} & 0 \\
0 & 0 & t_{zz}
\end{pmatrix} \end{math}
&
\begin{math} \begin{pmatrix}
t^\prime_{xx} & t^\prime_{xy} & 0 \\
-t^\prime_{xy} & t^\prime_{xx} & 0 \\
0 & 0 & t^\prime_{zz}
\end{pmatrix} \end{math}
&
\spacegroup{4/m}
&
\begin{math} \begin{pmatrix}
\sigma_{xx} & \sigma_{xy} & 0 \\
-\sigma_{xy} & \sigma_{xx} & 0 \\
0 & 0 & \sigma_{zz}
\end{pmatrix} \end{math}
&
\begin{math} \begin{pmatrix}
\sigma_{xx}^{z} & \sigma_{xy}^{z} & 0 \\
-\sigma_{xy}^{z} & \sigma_{xx}^{z} & 0 \\
0 & 0 & \sigma_{zz}^{z}
\end{pmatrix} \end{math}
\\[3ex]
\spacegroup{\bar 4}
&
\begin{math} \begin{pmatrix}
t_{xx} & t_{xy} & 0 \\
t_{xy} & -t_{xx} & 0 \\
0 & 0 & 0
\end{pmatrix} \end{math}
&
\begin{math} \begin{pmatrix}
t^\prime_{xx} & t^\prime_{xy} & 0 \\
t^\prime_{xy} & -t^\prime_{xx} & 0 \\
0 & 0 & 0
\end{pmatrix} \end{math}
&
\spacegroup{4/m}
&
\begin{math} \begin{pmatrix}
\sigma_{xx} & \sigma_{xy} & 0 \\
-\sigma_{xy} & \sigma_{xx} & 0 \\
0 & 0 & \sigma_{zz}
\end{pmatrix} \end{math}
&
\begin{math} \begin{pmatrix}
\sigma_{xx}^{z} & \sigma_{xy}^{z} & 0 \\
-\sigma_{xy}^{z} & \sigma_{xx}^{z} & 0 \\
0 & 0 & \sigma_{zz}^{z}
\end{pmatrix} \end{math}
\\[3ex]
\spacegroup{422}
&
\begin{math} \begin{pmatrix}
t_{xx} & 0 & 0 \\
0 & t_{xx} & 0 \\
0 & 0 & t_{zz}
\end{pmatrix} \end{math}
&
\begin{math} \begin{pmatrix}
t^\prime_{xx} & 0 & 0 \\
0 & t^\prime_{xx} & 0 \\
0 & 0 & t^\prime_{zz}
\end{pmatrix} \end{math}
&
\spacegroup{4/mmm}
&
\begin{math} \begin{pmatrix}
\sigma_{xx} & 0 & 0 \\
0 & \sigma_{xx} & 0 \\
0 & 0 & \sigma_{zz}
\end{pmatrix} \end{math}
&
\begin{math} \begin{pmatrix}
0 & \sigma_{xy}^{z} & 0 \\
-\sigma_{xy}^{z} & 0 & 0 \\
0 & 0 & 0
\end{pmatrix} \end{math}
\\[3ex]
\spacegroup{4mm}
&
\begin{math} \begin{pmatrix}
0 & t_{xy} & 0 \\
-t_{xy} & 0 & 0 \\
0 & 0 & 0
\end{pmatrix} \end{math}
&
\begin{math} \begin{pmatrix}
0 & t^\prime_{xy} & 0 \\
-t^\prime_{xy} & 0 & 0 \\
0 & 0 & 0
\end{pmatrix} \end{math}
&
\spacegroup{4/mmm}
&
\begin{math} \begin{pmatrix}
\sigma_{xx} & 0 & 0 \\
0 & \sigma_{xx} & 0 \\
0 & 0 & \sigma_{zz}
\end{pmatrix} \end{math}
&
\begin{math} \begin{pmatrix}
0 & \sigma_{xy}^{z} & 0 \\
-\sigma_{xy}^{z} & 0 & 0 \\
0 & 0 & 0
\end{pmatrix} \end{math}
\\[3ex]
\spacegroup{\bar 42m}
&
\begin{math} \begin{pmatrix}
t_{xx} & 0 & 0 \\
0 & -t_{xx} & 0 \\
0 & 0 & 0
\end{pmatrix} \end{math}
&
\begin{math} \begin{pmatrix}
t^\prime_{xx} & 0 & 0 \\
0 & -t^\prime_{xx} & 0 \\
0 & 0 & 0
\end{pmatrix} \end{math}
&
\spacegroup{4/mmm}
&
\begin{math} \begin{pmatrix}
\sigma_{xx} & 0 & 0 \\
0 & \sigma_{xx} & 0 \\
0 & 0 & \sigma_{zz}
\end{pmatrix} \end{math}
&
\begin{math} \begin{pmatrix}
0 & \sigma_{xy}^{z} & 0 \\
-\sigma_{xy}^{z} & 0 & 0 \\
0 & 0 & 0
\end{pmatrix} \end{math}
\\[3ex]
\spacegroup{3}
&
\begin{math} \begin{pmatrix}
t_{xx} & t_{xy} & 0 \\
-t_{xy} & t_{xx} & 0 \\
0 & 0 & t_{zz}
\end{pmatrix} \end{math}
&
\begin{math} \begin{pmatrix}
t^\prime_{xx} & t^\prime_{xy} & 0 \\
-t^\prime_{xy} & t^\prime_{xx} & 0 \\
0 & 0 & t^\prime_{zz}
\end{pmatrix} \end{math}
&
\spacegroup{\bar 3}
&
\begin{math} \begin{pmatrix}
\sigma_{xx} & \sigma_{xy} & 0 \\
-\sigma_{xy} & \sigma_{xx} & 0 \\
0 & 0 & \sigma_{zz}
\end{pmatrix} \end{math}
&
\begin{math} \begin{pmatrix}
\sigma_{xx}^{z} & \sigma_{xy}^{z} & 0 \\
-\sigma_{xy}^{z} & \sigma_{xx}^{z} & 0 \\
0 & 0 & \sigma_{zz}^{z}
\end{pmatrix} \end{math}
\\[3ex]
\spacegroup{312}
&
\begin{math} \begin{pmatrix}
t_{xx} & 0 & 0 \\
0 & t_{xx} & 0 \\
0 & 0 & t_{zz}
\end{pmatrix} \end{math}
&
\begin{math} \begin{pmatrix}
t^\prime_{xx} & 0 & 0 \\
0 & t^\prime_{xx} & 0 \\
0 & 0 & t^\prime_{zz}
\end{pmatrix} \end{math}
&
\spacegroup{\bar 31m}
&
\begin{math} \begin{pmatrix}
\sigma_{xx} & 0 & 0 \\
0 & \sigma_{xx} & 0 \\
0 & 0 & \sigma_{zz}
\end{pmatrix} \end{math}
&
\begin{math} \begin{pmatrix}
0 & \sigma_{xy}^{z} & 0 \\
-\sigma_{xy}^{z} & 0 & 0 \\
0 & 0 & 0
\end{pmatrix} \end{math}
\\[3ex]
\spacegroup{31m}
&
\begin{math} \begin{pmatrix}
0 & t_{xy} & 0 \\
-t_{xy} & 0 & 0 \\
0 & 0 & 0
\end{pmatrix} \end{math}
&
\begin{math} \begin{pmatrix}
0 & t^\prime_{xy} & 0 \\
-t^\prime_{xy} & 0 & 0 \\
0 & 0 & 0
\end{pmatrix} \end{math}
&
\spacegroup{\bar 31m}
&
\begin{math} \begin{pmatrix}
\sigma_{xx} & 0 & 0 \\
0 & \sigma_{xx} & 0 \\
0 & 0 & \sigma_{zz}
\end{pmatrix} \end{math}
&
\begin{math} \begin{pmatrix}
0 & \sigma_{xy}^{z} & 0 \\
-\sigma_{xy}^{z} & 0 & 0 \\
0 & 0 & 0
\end{pmatrix} \end{math}
\\[3ex]
\spacegroup{6}
&
\begin{math} \begin{pmatrix}
t_{xx} & t_{xy} & 0 \\
-t_{xy} & t_{xx} & 0 \\
0 & 0 & t_{zz}
\end{pmatrix} \end{math}
&
\begin{math} \begin{pmatrix}
t^\prime_{xx} & t^\prime_{xy} & 0 \\
-t^\prime_{xy} & t^\prime_{xx} & 0 \\
0 & 0 & t^\prime_{zz}
\end{pmatrix} \end{math}
&
\spacegroup{6/m}
&
\begin{math} \begin{pmatrix}
\sigma_{xx} & \sigma_{xy} & 0 \\
-\sigma_{xy} & \sigma_{xx} & 0 \\
0 & 0 & \sigma_{zz}
\end{pmatrix} \end{math}
&
\begin{math} \begin{pmatrix}
\sigma_{xx}^{z} & \sigma_{xy}^{z} & 0 \\
-\sigma_{xy}^{z} & \sigma_{xx}^{z} & 0 \\
0 & 0 & \sigma_{zz}^{z}
\end{pmatrix} \end{math}
\\[3ex]
\spacegroup{622}
&
\begin{math} \begin{pmatrix}
t_{xx} & 0 & 0 \\
0 & t_{xx} & 0 \\
0 & 0 & t_{zz}
\end{pmatrix} \end{math}
&
\begin{math} \begin{pmatrix}
t^\prime_{xx} & 0 & 0 \\
0 & t^\prime_{xx} & 0 \\
0 & 0 & t^\prime_{zz}
\end{pmatrix} \end{math}
&
\spacegroup{6/mmm}
&
\begin{math} \begin{pmatrix}
\sigma_{xx} & 0 & 0 \\
0 & \sigma_{xx} & 0 \\
0 & 0 & \sigma_{zz}
\end{pmatrix} \end{math}
&
\begin{math} \begin{pmatrix}
0 & \sigma_{xy}^{z} & 0 \\
-\sigma_{xy}^{z} & 0 & 0 \\
0 & 0 & 0
\end{pmatrix} \end{math}
\\[3ex]
\spacegroup{6mm}
&
\begin{math} \begin{pmatrix}
0 & t_{xy} & 0 \\
-t_{xy} & 0 & 0 \\
0 & 0 & 0
\end{pmatrix} \end{math}
&
\begin{math} \begin{pmatrix}
0 & t^\prime_{xy} & 0 \\
-t^\prime_{xy} & 0 & 0 \\
0 & 0 & 0
\end{pmatrix} \end{math}
&
\spacegroup{6/mmm}
&
\begin{math} \begin{pmatrix}
\sigma_{xx} & 0 & 0 \\
0 & \sigma_{xx} & 0 \\
0 & 0 & \sigma_{zz}
\end{pmatrix} \end{math}
&
\begin{math} \begin{pmatrix}
0 & \sigma_{xy}^{z} & 0 \\
-\sigma_{xy}^{z} & 0 & 0 \\
0 & 0 & 0
\end{pmatrix} \end{math}
\\[3ex]
\spacegroup{23}
&
\begin{math} \begin{pmatrix}
t_{xx} & 0 & 0 \\
0 & t_{xx} & 0 \\
0 & 0 & t_{xx}
\end{pmatrix} \end{math}
&
\begin{math} \begin{pmatrix}
t^\prime_{xx} & 0 & 0 \\
0 & t^\prime_{xx} & 0 \\
0 & 0 & t^\prime_{xx}
\end{pmatrix} \end{math}
&
\spacegroup{m\bar 3}
&
\begin{math} \begin{pmatrix}
\sigma_{xx} & 0 & 0 \\
0 & \sigma_{xx} & 0 \\
0 & 0 & \sigma_{xx}
\end{pmatrix} \end{math}
&
\begin{math} \begin{pmatrix}
0 & \sigma_{xy}^{z} & 0 \\
xzy & 0 & 0 \\
0 & 0 & 0
\end{pmatrix} \end{math}
\\[3ex]
\spacegroup{432}
&
\begin{math} \begin{pmatrix}
t_{xx} & 0 & 0 \\
0 & t_{xx} & 0 \\
0 & 0 & t_{xx}
\end{pmatrix} \end{math}
&
\begin{math} \begin{pmatrix}
t^\prime_{xx} & 0 & 0 \\
0 & t^\prime_{xx} & 0 \\
0 & 0 & t^\prime_{xx}
\end{pmatrix} \end{math}
&
\spacegroup{m\bar 3m}
&
\begin{math} \begin{pmatrix}
\sigma_{xx} & 0 & 0 \\
0 & \sigma_{xx} & 0 \\
0 & 0 & \sigma_{xx}
\end{pmatrix} \end{math}
&
\begin{math} \begin{pmatrix}
0 & \sigma_{xy}^{z} & 0 \\
-\sigma_{xy}^{z} & 0 & 0 \\
0 & 0 & 0
\end{pmatrix} \end{math}
\\[3ex]
\end{tabular}

 \caption{
\label{TAB-TORTENb}
Shape of the direct and inverse torkance tensors, \TORTEN\ and
\TORTEN$^\prime$, for all magnetic point groups of category b). Note that since
these do not contain time-reversal, neither as an element on its own nor in
combination with a spatial operation, the two tensors are unconnected and
identical in shape. The third and fourth columns show the electrical
conductivity tensor \SIGTEN\ and the spin conductivity tensor \SIGTEN$^k$ for polarization
along the principal axis $k$, respectively, for the corresponding
magnetic Laue groups. See Ref.~\onlinecite{SKWE15} for the two remaining
polarization directions and further details on conventions and notation.}
\end{table*}
%
%
%
\begin{table*}
  \tabulinesep\matrixtablelinesep
  \renewcommand{\arraystretch}{\matrixtablearraystretch}
  \setlength\arraycolsep{\matrixtablearraycolsep}
  \centering
\begin{tabular}{>{\centering}m{5.0cm}cccccc}
\tableheading{magnetic point group} & \tableheading{\TORTEN} & \tableheading{\TORTEN$^\prime$} & \tableheading{magnetic Laue group} & \tableheading{\SIGTEN} & \tableheading{\SIGTEN$^k$} \\[4mm]
\spacegroup{\bar 1'}
&
\begin{math} \begin{pmatrix}
t_{xx} & t_{xy} & t_{xz} \\
t_{yx} & t_{yy} & t_{yz} \\
t_{zx} & t_{zy} & t_{zz}
\end{pmatrix} \end{math}
&
\begin{math} \begin{pmatrix}
t_{xx} & t_{yx} & t_{zx} \\
t_{xy} & t_{yy} & t_{zy} \\
t_{xz} & t_{yz} & t_{zz}
\end{pmatrix} \end{math}
&
\spacegroup{\bar 11'}
&
\begin{math} \begin{pmatrix}
\sigma_{xx} & \sigma_{xy} & \sigma_{xz} \\
\sigma_{xy} & \sigma_{yy} & \sigma_{yz} \\
\sigma_{xz} & \sigma_{yz} & \sigma_{zz}
\end{pmatrix} \end{math}
&
\begin{math} \begin{pmatrix}
\sigma_{xx}^{z} & \sigma_{xy}^{z} & \sigma_{xz}^{z} \\
\sigma_{yx}^{z} & \sigma_{yy}^{z} & \sigma_{yz}^{z} \\
\sigma_{zx}^{z} & \sigma_{zy}^{z} & \sigma_{zz}^{z}
\end{pmatrix} \end{math}
\\[3ex]
\spacegroup{2'}
&
\begin{math} \begin{pmatrix}
t_{xx} & t_{xy} & t_{xz} \\
t_{yx} & t_{yy} & t_{yz} \\
t_{zx} & t_{zy} & t_{zz}
\end{pmatrix} \end{math}
&
\begin{math} \begin{pmatrix}
-t_{xx} & t_{yx} & -t_{zx} \\
t_{xy} & -t_{yy} & t_{zy} \\
-t_{xz} & t_{yz} & -t_{zz}
\end{pmatrix} \end{math}
&
\spacegroup{2'/m'}
&
\begin{math} \begin{pmatrix}
\sigma_{xx} & \sigma_{xy} & \sigma_{xz} \\
-\sigma_{xy} & \sigma_{yy} & \sigma_{yz} \\
\sigma_{xz} & -\sigma_{yz} & \sigma_{zz}
\end{pmatrix} \end{math}
&
\begin{math} \begin{pmatrix}
\sigma_{xx}^{y} & \sigma_{xy}^{y} & \sigma_{xz}^{y} \\
\sigma_{yx}^{y} & \sigma_{yy}^{y} & \sigma_{yz}^{y} \\
\sigma_{zx}^{y} & \sigma_{zy}^{y} & \sigma_{zz}^{y}
\end{pmatrix} \end{math}
\\[3ex]
\spacegroup{m'}
&
\begin{math} \begin{pmatrix}
t_{xx} & t_{xy} & t_{xz} \\
t_{yx} & t_{yy} & t_{yz} \\
t_{zx} & t_{zy} & t_{zz}
\end{pmatrix} \end{math}
&
\begin{math} \begin{pmatrix}
t_{xx} & -t_{yx} & t_{zx} \\
-t_{xy} & t_{yy} & -t_{zy} \\
t_{xz} & -t_{yz} & t_{zz}
\end{pmatrix} \end{math}
&
\spacegroup{2'/m'}
&
\begin{math} \begin{pmatrix}
\sigma_{xx} & \sigma_{xy} & \sigma_{xz} \\
-\sigma_{xy} & \sigma_{yy} & \sigma_{yz} \\
\sigma_{xz} & -\sigma_{yz} & \sigma_{zz}
\end{pmatrix} \end{math}
&
\begin{math} \begin{pmatrix}
\sigma_{xx}^{y} & \sigma_{xy}^{y} & \sigma_{xz}^{y} \\
\sigma_{yx}^{y} & \sigma_{yy}^{y} & \sigma_{yz}^{y} \\
\sigma_{zx}^{y} & \sigma_{zy}^{y} & \sigma_{zz}^{y}
\end{pmatrix} \end{math}
\\[3ex]
\spacegroup{2/m'}
&
\begin{math} \begin{pmatrix}
t_{xx} & 0 & t_{xz} \\
0 & t_{yy} & 0 \\
t_{zx} & 0 & t_{zz}
\end{pmatrix} \end{math}
&
\begin{math} \begin{pmatrix}
t_{xx} & 0 & t_{zx} \\
0 & t_{yy} & 0 \\
t_{xz} & 0 & t_{zz}
\end{pmatrix} \end{math}
&
\spacegroup{2/m1'}
&
\begin{math} \begin{pmatrix}
\sigma_{xx} & 0 & \sigma_{xz} \\
0 & \sigma_{yy} & 0 \\
\sigma_{xz} & 0 & \sigma_{zz}
\end{pmatrix} \end{math}
&
\begin{math} \begin{pmatrix}
\sigma_{xx}^{y} & 0 & \sigma_{xz}^{y} \\
0 & \sigma_{yy}^{y} & 0 \\
\sigma_{zx}^{y} & 0 & \sigma_{zz}^{y}
\end{pmatrix} \end{math}
\\[3ex]
\spacegroup{2'/m}
&
\begin{math} \begin{pmatrix}
0 & t_{xy} & 0 \\
t_{yx} & 0 & t_{yz} \\
0 & t_{zy} & 0
\end{pmatrix} \end{math}
&
\begin{math} \begin{pmatrix}
0 & t_{yx} & 0 \\
t_{xy} & 0 & t_{zy} \\
0 & t_{yz} & 0
\end{pmatrix} \end{math}
&
\spacegroup{2/m1'}
&
\begin{math} \begin{pmatrix}
\sigma_{xx} & 0 & \sigma_{xz} \\
0 & \sigma_{yy} & 0 \\
\sigma_{xz} & 0 & \sigma_{zz}
\end{pmatrix} \end{math}
&
\begin{math} \begin{pmatrix}
\sigma_{xx}^{y} & 0 & \sigma_{xz}^{y} \\
0 & \sigma_{yy}^{y} & 0 \\
\sigma_{zx}^{y} & 0 & \sigma_{zz}^{y}
\end{pmatrix} \end{math}
\\[3ex]
\spacegroup{2'2'2}
&
\begin{math} \begin{pmatrix}
t_{xx} & t_{xy} & 0 \\
t_{yx} & t_{yy} & 0 \\
0 & 0 & t_{zz}
\end{pmatrix} \end{math}
&
\begin{math} \begin{pmatrix}
-t_{xx} & t_{yx} & 0 \\
t_{xy} & -t_{yy} & 0 \\
0 & 0 & -t_{zz}
\end{pmatrix} \end{math}
&
\spacegroup{m'm'm}
&
\begin{math} \begin{pmatrix}
\sigma_{xx} & \sigma_{xy} & 0 \\
-\sigma_{xy} & \sigma_{yy} & 0 \\
0 & 0 & \sigma_{zz}
\end{pmatrix} \end{math}
&
\begin{math} \begin{pmatrix}
\sigma_{xx}^{z} & \sigma_{xy}^{z} & 0 \\
\sigma_{yx}^{z} & \sigma_{yy}^{z} & 0 \\
0 & 0 & \sigma_{zz}^{z}
\end{pmatrix} \end{math}
\\[3ex]
\spacegroup{m'm'2}
&
\begin{math} \begin{pmatrix}
t_{xx} & t_{xy} & 0 \\
t_{yx} & t_{yy} & 0 \\
0 & 0 & t_{zz}
\end{pmatrix} \end{math}
&
\begin{math} \begin{pmatrix}
t_{xx} & -t_{yx} & 0 \\
-t_{xy} & t_{yy} & 0 \\
0 & 0 & t_{zz}
\end{pmatrix} \end{math}
&
\spacegroup{m'm'm}
&
\begin{math} \begin{pmatrix}
\sigma_{xx} & \sigma_{xy} & 0 \\
-\sigma_{xy} & \sigma_{yy} & 0 \\
0 & 0 & \sigma_{zz}
\end{pmatrix} \end{math}
&
\begin{math} \begin{pmatrix}
\sigma_{xx}^{z} & \sigma_{xy}^{z} & 0 \\
\sigma_{yx}^{z} & \sigma_{yy}^{z} & 0 \\
0 & 0 & \sigma_{zz}^{z}
\end{pmatrix} \end{math}
\\[3ex]
\spacegroup{m'm2'}
&
\begin{math} \begin{pmatrix}
0 & t_{xy} & 0 \\
t_{yx} & 0 & t_{yz} \\
0 & t_{zy} & 0
\end{pmatrix} \end{math}
&
\begin{math} \begin{pmatrix}
0 & -t_{yx} & 0 \\
-t_{xy} & 0 & t_{zy} \\
0 & t_{yz} & 0
\end{pmatrix} \end{math}
&
\spacegroup{m'm'm}
&
\begin{math} \begin{pmatrix}
\sigma_{xx} & 0 & \sigma_{xz} \\
0 & \sigma_{yy} & 0 \\
-\sigma_{xz} & 0 & \sigma_{zz}
\end{pmatrix} \end{math}
&
\begin{math} \begin{pmatrix}
0 & \sigma_{xy}^{z} & 0 \\
\sigma_{yx}^{z} & 0 & \sigma_{yz}^{z} \\
0 & \sigma_{zy}^{z} & 0
\end{pmatrix} \end{math}
\\[3ex]
\spacegroup{m'm'm'}
&
\begin{math} \begin{pmatrix}
t_{xx} & 0 & 0 \\
0 & t_{yy} & 0 \\
0 & 0 & t_{zz}
\end{pmatrix} \end{math}
&
\begin{math} \begin{pmatrix}
t_{xx} & 0 & 0 \\
0 & t_{yy} & 0 \\
0 & 0 & t_{zz}
\end{pmatrix} \end{math}
&
\spacegroup{mmm1'}
&
\begin{math} \begin{pmatrix}
\sigma_{xx} & 0 & 0 \\
0 & \sigma_{yy} & 0 \\
0 & 0 & \sigma_{zz}
\end{pmatrix} \end{math}
&
\begin{math} \begin{pmatrix}
0 & \sigma_{xy}^{z} & 0 \\
\sigma_{yx}^{z} & 0 & 0 \\
0 & 0 & 0
\end{pmatrix} \end{math}
\\[3ex]
\spacegroup{m'mm}
&
\begin{math} \begin{pmatrix}
0 & 0 & 0 \\
0 & 0 & t_{yz} \\
0 & t_{zy} & 0
\end{pmatrix} \end{math}
&
\begin{math} \begin{pmatrix}
0 & 0 & 0 \\
0 & 0 & t_{zy} \\
0 & t_{yz} & 0
\end{pmatrix} \end{math}
&
\spacegroup{mmm1'}
&
\begin{math} \begin{pmatrix}
\sigma_{xx} & 0 & 0 \\
0 & \sigma_{yy} & 0 \\
0 & 0 & \sigma_{zz}
\end{pmatrix} \end{math}
&
\begin{math} \begin{pmatrix}
0 & \sigma_{xy}^{z} & 0 \\
\sigma_{yx}^{z} & 0 & 0 \\
0 & 0 & 0
\end{pmatrix} \end{math}
\\[3ex]
\end{tabular}

 \caption{
\label{TAB-TORTENc}
Shape of the direct and inverse torkance tensors, \TORTEN\ and
\TORTEN$^\prime$, for magnetic point groups of category c). Note that the two
tensors usually differ in shape, depending on which spatial operation is
combined with time-reversal. The third and fourth columns show the electrical
conductivity tensor \SIGTEN\ and the spin conductivity tensor \SIGTEN$^k$ for
polarization along the principal axis $k$, respectively, for the corresponding
magnetic Laue groups. See Ref.~\onlinecite{SKWE15} for the two remaining
polarization directions and further details on conventions and notation. This
Table contains only groups with a principal axis of order $O(k) \le 2$
and is continued in Tables~\ref{TAB-TORTENc1b}--\ref{TAB-TORTENc2c}.}
\end{table*}
%
\sw{Naturally,} this excludes all magnetic point groups \sw{of category a)}
corresponding to a non-magnetic solid,
\sw{
i.e., that contain time-reversal as separate element.\cite{Kle66,SKWE15}
}

\sw{
Comparing the results for magnetic point groups of categories b) in
Table~\ref{TAB-TORTENb} and c) in
Tables~\ref{TAB-TORTENc}--\ref{TAB-TORTENc2c}, one notes that those of the
former exhibit identical direct and inverse torkance tensor shapes, \TORTEN\
and \TORTEN$^\prime$, while for those of the latter the two tensors usually
differ in shape but nevertheless are connected to each other. This becomes
obvious when looking at Eq.~\eqref{sym-anti-unitary}: if there was
time-reversal as a separate operation, as in a group of category a), the
corresponding spatial operation $R$ would be the identity, and therefore
$t_{\mu\nu} = -t^\prime_{\nu\mu}$ for all tensor elements, i.e., something
quite similar to the usual Onsager relations would hold. When there are no
time-reversal-connected, i.e., anti-unitary operations in the group, as in
category b), Eq.~\eqref{sym-anti-unitary} does not apply at all and the shape
of \TORTEN$^\prime$\ is given exclusively by Eq.~\eqref{sym-unitary} and
therefore identical to that of \TORTEN. For the magnetic point groups of
category c), where time-reversal appears only in connection with a spatial
operation $R$, the shape of \TORTEN$^\prime$\ is determined by $\matrg{D}(R)$,
i.e., the nature of the operation connecting \TORTEN\ and \TORTEN$^\prime$.
}

In addition one notices that none of the magnetic point groups listed in
Tables~\ref{TAB-TORTENb}--\ref{TAB-TORTENc2c} contains the spatial inversion as
an element.  This central restriction -- missing inversion symmetry -- has been
pointed out  before by Manchon and Zhang  \cite{MZ09} as well as Garate and
MacDonald \cite{GM09a} on the basis of restricted  model considerations.
This basic requirement is explained here on group-theoretical grounds by the
transformation properties of the operators appearing in the linear response
expression. The torque operator, represented by the vector product of
magnetization and effective magnetic field -- both pseudo vectors symmetric
under spatial inversion but anti-symmetric under time reversal -- hence
transforms as time-reversal symmetric pseudo vector, while the electric current
density operator as a proper vector is anti-symmetric under both. Therefore the
product of the two is both time-reversal- and inversion anti-symmetric.
Correspondingly, the shapes of direct and inverse torkance tensors are determined
by the magnetic point group of a solid, in contrast for example to the electrical
conductivity and thermoelectric tensors,\cite{Kle66} as well as to the spin
conductivity tensor.\cite{SKWE15} 

Since the operators for electric and heat current densities transform identical
under all space-time symmetry operations relevant for
solids,\cite{Kle66,SKWE15} the tensor shapes will stay unaltered when the
electric field is replaced by a temperature gradient. In other words, the
shapes given here apply also for the direct and inverse thermal spin-orbit
torque effect discussed recently by G\'eranton \ea\cite{GFBM15} and Freimuth
\ea\cite{FBM16}.

Finally, it should be mentioned that the results for the torkance tensors
presented in Tables~\ref{TAB-TORTENb}--\ref{TAB-TORTENc2c} have been
independently checked for a number of systems, including \spacegroup{m'm'm},
\spacegroup{4/mm'm'}, \spacegroup{6/mm'm'}, \spacegroup{\bar 3m'} with
vanishing torkance and \spacegroup{1}, \spacegroup{m'}, \spacegroup{m'm'2},
\spacegroup{\bar 42'm'}, \spacegroup{3m'}, \spacegroup{\bar 6'2m'},
\spacegroup{6/m'm'm'} with finite torkance, by numerical calculations using the
implementation described above.
%
%
%
%
\begin{table*}
  \tabulinesep\matrixtablelinesep
  \renewcommand{\arraystretch}{\matrixtablearraystretch}
  \setlength\arraycolsep{\matrixtablearraycolsep}
  \centering
\begin{tabular}{>{\centering}m{5.0cm}cccccc}
\tableheading{magnetic point group} & \tableheading{\TORTEN} & \tableheading{\TORTEN$^\prime$} & \tableheading{magnetic Laue group} & \tableheading{\SIGTEN} & \tableheading{\SIGTEN$^k$} \\[4mm]
\spacegroup{4'}
&
\begin{math} \begin{pmatrix}
t_{xx} & t_{xy} & 0 \\
t_{yx} & t_{yy} & 0 \\
0 & 0 & t_{zz}
\end{pmatrix} \end{math}
&
\begin{math} \begin{pmatrix}
-t_{yy} & t_{xy} & 0 \\
t_{yx} & -t_{xx} & 0 \\
0 & 0 & -t_{zz}
\end{pmatrix} \end{math}
&
\spacegroup{4'/m}
&
\begin{math} \begin{pmatrix}
\sigma_{xx} & 0 & 0 \\
0 & \sigma_{xx} & 0 \\
0 & 0 & \sigma_{zz}
\end{pmatrix} \end{math}
&
\begin{math} \begin{pmatrix}
\sigma_{xx}^{z} & \sigma_{xy}^{z} & 0 \\
\sigma_{yx}^{z} & \sigma_{yy}^{z} & 0 \\
0 & 0 & \sigma_{zz}^{z}
\end{pmatrix} \end{math}
\\[3ex]
\spacegroup{\bar 4'}
&
\begin{math} \begin{pmatrix}
t_{xx} & t_{xy} & 0 \\
t_{yx} & t_{yy} & 0 \\
0 & 0 & t_{zz}
\end{pmatrix} \end{math}
&
\begin{math} \begin{pmatrix}
t_{yy} & -t_{xy} & 0 \\
-t_{yx} & t_{xx} & 0 \\
0 & 0 & t_{zz}
\end{pmatrix} \end{math}
&
\spacegroup{4'/m}
&
\begin{math} \begin{pmatrix}
\sigma_{xx} & 0 & 0 \\
0 & \sigma_{xx} & 0 \\
0 & 0 & \sigma_{zz}
\end{pmatrix} \end{math}
&
\begin{math} \begin{pmatrix}
\sigma_{xx}^{z} & \sigma_{xy}^{z} & 0 \\
\sigma_{yx}^{z} & \sigma_{yy}^{z} & 0 \\
0 & 0 & \sigma_{zz}^{z}
\end{pmatrix} \end{math}
\\[3ex]
\spacegroup{4/m'}
&
\begin{math} \begin{pmatrix}
t_{xx} & t_{xy} & 0 \\
-t_{xy} & t_{xx} & 0 \\
0 & 0 & t_{zz}
\end{pmatrix} \end{math}
&
\begin{math} \begin{pmatrix}
t_{xx} & -t_{xy} & 0 \\
t_{xy} & t_{xx} & 0 \\
0 & 0 & t_{zz}
\end{pmatrix} \end{math}
&
\spacegroup{4/m1'}
&
\begin{math} \begin{pmatrix}
\sigma_{xx} & 0 & 0 \\
0 & \sigma_{xx} & 0 \\
0 & 0 & \sigma_{zz}
\end{pmatrix} \end{math}
&
\begin{math} \begin{pmatrix}
\sigma_{xx}^{z} & \sigma_{xy}^{z} & 0 \\
-\sigma_{xy}^{z} & \sigma_{xx}^{z} & 0 \\
0 & 0 & \sigma_{zz}^{z}
\end{pmatrix} \end{math}
\\[3ex]
\spacegroup{4'/m'}
&
\begin{math} \begin{pmatrix}
t_{xx} & t_{xy} & 0 \\
t_{xy} & -t_{xx} & 0 \\
0 & 0 & 0
\end{pmatrix} \end{math}
&
\begin{math} \begin{pmatrix}
t_{xx} & t_{xy} & 0 \\
t_{xy} & -t_{xx} & 0 \\
0 & 0 & 0
\end{pmatrix} \end{math}
&
\spacegroup{4/m1'}
&
\begin{math} \begin{pmatrix}
\sigma_{xx} & 0 & 0 \\
0 & \sigma_{xx} & 0 \\
0 & 0 & \sigma_{zz}
\end{pmatrix} \end{math}
&
\begin{math} \begin{pmatrix}
\sigma_{xx}^{z} & \sigma_{xy}^{z} & 0 \\
-\sigma_{xy}^{z} & \sigma_{xx}^{z} & 0 \\
0 & 0 & \sigma_{zz}^{z}
\end{pmatrix} \end{math}
\\[3ex]
\spacegroup{4'22'}
&
\begin{math} \begin{pmatrix}
t_{xx} & 0 & 0 \\
0 & t_{yy} & 0 \\
0 & 0 & t_{zz}
\end{pmatrix} \end{math}
&
\begin{math} \begin{pmatrix}
-t_{yy} & 0 & 0 \\
0 & -t_{xx} & 0 \\
0 & 0 & -t_{zz}
\end{pmatrix} \end{math}
&
\spacegroup{4'/mmm'}
&
\begin{math} \begin{pmatrix}
\sigma_{xx} & 0 & 0 \\
0 & \sigma_{xx} & 0 \\
0 & 0 & \sigma_{zz}
\end{pmatrix} \end{math}
&
\begin{math} \begin{pmatrix}
0 & \sigma_{xy}^{z} & 0 \\
\sigma_{yx}^{z} & 0 & 0 \\
0 & 0 & 0
\end{pmatrix} \end{math}
\\[3ex]
\spacegroup{42'2'}
&
\begin{math} \begin{pmatrix}
t_{xx} & t_{xy} & 0 \\
-t_{xy} & t_{xx} & 0 \\
0 & 0 & t_{zz}
\end{pmatrix} \end{math}
&
\begin{math} \begin{pmatrix}
-t_{xx} & -t_{xy} & 0 \\
t_{xy} & -t_{xx} & 0 \\
0 & 0 & -t_{zz}
\end{pmatrix} \end{math}
&
\spacegroup{4/mm'm'}
&
\begin{math} \begin{pmatrix}
\sigma_{xx} & \sigma_{xy} & 0 \\
-\sigma_{xy} & \sigma_{xx} & 0 \\
0 & 0 & \sigma_{zz}
\end{pmatrix} \end{math}
&
\begin{math} \begin{pmatrix}
\sigma_{xx}^{z} & \sigma_{xy}^{z} & 0 \\
-\sigma_{xy}^{z} & \sigma_{xx}^{z} & 0 \\
0 & 0 & \sigma_{zz}^{z}
\end{pmatrix} \end{math}
\\[3ex]
\spacegroup{4'mm'}
&
\begin{math} \begin{pmatrix}
0 & t_{xy} & 0 \\
t_{yx} & 0 & 0 \\
0 & 0 & 0
\end{pmatrix} \end{math}
&
\begin{math} \begin{pmatrix}
0 & t_{xy} & 0 \\
t_{yx} & 0 & 0 \\
0 & 0 & 0
\end{pmatrix} \end{math}
&
\spacegroup{4'/mmm'}
&
\begin{math} \begin{pmatrix}
\sigma_{xx} & 0 & 0 \\
0 & \sigma_{xx} & 0 \\
0 & 0 & \sigma_{zz}
\end{pmatrix} \end{math}
&
\begin{math} \begin{pmatrix}
0 & \sigma_{xy}^{z} & 0 \\
\sigma_{yx}^{z} & 0 & 0 \\
0 & 0 & 0
\end{pmatrix} \end{math}
\\[3ex]
\spacegroup{4m'm'}
&
\begin{math} \begin{pmatrix}
t_{xx} & t_{xy} & 0 \\
-t_{xy} & t_{xx} & 0 \\
0 & 0 & t_{zz}
\end{pmatrix} \end{math}
&
\begin{math} \begin{pmatrix}
t_{xx} & t_{xy} & 0 \\
-t_{xy} & t_{xx} & 0 \\
0 & 0 & t_{zz}
\end{pmatrix} \end{math}
&
\spacegroup{4/mm'm'}
&
\begin{math} \begin{pmatrix}
\sigma_{xx} & \sigma_{xy} & 0 \\
-\sigma_{xy} & \sigma_{xx} & 0 \\
0 & 0 & \sigma_{zz}
\end{pmatrix} \end{math}
&
\begin{math} \begin{pmatrix}
\sigma_{xx}^{z} & \sigma_{xy}^{z} & 0 \\
-\sigma_{xy}^{z} & \sigma_{xx}^{z} & 0 \\
0 & 0 & \sigma_{zz}^{z}
\end{pmatrix} \end{math}
\\[3ex]
\spacegroup{\bar 4'2m'}
&
\begin{math} \begin{pmatrix}
t_{xx} & 0 & 0 \\
0 & t_{yy} & 0 \\
0 & 0 & t_{zz}
\end{pmatrix} \end{math}
&
\begin{math} \begin{pmatrix}
t_{yy} & 0 & 0 \\
0 & t_{xx} & 0 \\
0 & 0 & t_{zz}
\end{pmatrix} \end{math}
&
\spacegroup{4'/mmm'}
&
\begin{math} \begin{pmatrix}
\sigma_{xx} & 0 & 0 \\
0 & \sigma_{xx} & 0 \\
0 & 0 & \sigma_{zz}
\end{pmatrix} \end{math}
&
\begin{math} \begin{pmatrix}
0 & \sigma_{xy}^{z} & 0 \\
\sigma_{yx}^{z} & 0 & 0 \\
0 & 0 & 0
\end{pmatrix} \end{math}
\\[3ex]
\spacegroup{\bar 4'm2'}
&
\begin{math} \begin{pmatrix}
0 & t_{xy} & 0 \\
t_{yx} & 0 & 0 \\
0 & 0 & 0
\end{pmatrix} \end{math}
&
\begin{math} \begin{pmatrix}
0 & -t_{xy} & 0 \\
-t_{yx} & 0 & 0 \\
0 & 0 & 0
\end{pmatrix} \end{math}
&
\spacegroup{4'/mmm'}
&
\begin{math} \begin{pmatrix}
\sigma_{xx} & 0 & 0 \\
0 & \sigma_{xx} & 0 \\
0 & 0 & \sigma_{zz}
\end{pmatrix} \end{math}
&
\begin{math} \begin{pmatrix}
0 & \sigma_{xy}^{z} & 0 \\
\sigma_{yx}^{z} & 0 & 0 \\
0 & 0 & 0
\end{pmatrix} \end{math}
\\[3ex]
\spacegroup{\bar 42'm'}
&
\begin{math} \begin{pmatrix}
t_{xx} & t_{xy} & 0 \\
t_{xy} & -t_{xx} & 0 \\
0 & 0 & 0
\end{pmatrix} \end{math}
&
\begin{math} \begin{pmatrix}
-t_{xx} & t_{xy} & 0 \\
t_{xy} & t_{xx} & 0 \\
0 & 0 & 0
\end{pmatrix} \end{math}
&
\spacegroup{4/mm'm'}
&
\begin{math} \begin{pmatrix}
\sigma_{xx} & \sigma_{xy} & 0 \\
-\sigma_{xy} & \sigma_{xx} & 0 \\
0 & 0 & \sigma_{zz}
\end{pmatrix} \end{math}
&
\begin{math} \begin{pmatrix}
\sigma_{xx}^{z} & \sigma_{xy}^{z} & 0 \\
-\sigma_{xy}^{z} & \sigma_{xx}^{z} & 0 \\
0 & 0 & \sigma_{zz}^{z}
\end{pmatrix} \end{math}
\\[3ex]
\spacegroup{4/m'm'm'}
&
\begin{math} \begin{pmatrix}
t_{xx} & 0 & 0 \\
0 & t_{xx} & 0 \\
0 & 0 & t_{zz}
\end{pmatrix} \end{math}
&
\begin{math} \begin{pmatrix}
t_{xx} & 0 & 0 \\
0 & t_{xx} & 0 \\
0 & 0 & t_{zz}
\end{pmatrix} \end{math}
&
\spacegroup{4/mmm1'}
&
\begin{math} \begin{pmatrix}
\sigma_{xx} & 0 & 0 \\
0 & \sigma_{xx} & 0 \\
0 & 0 & \sigma_{zz}
\end{pmatrix} \end{math}
&
\begin{math} \begin{pmatrix}
0 & \sigma_{xy}^{z} & 0 \\
-\sigma_{xy}^{z} & 0 & 0 \\
0 & 0 & 0
\end{pmatrix} \end{math}
\\[3ex]
\spacegroup{4/m'mm}
&
\begin{math} \begin{pmatrix}
0 & t_{xy} & 0 \\
-t_{xy} & 0 & 0 \\
0 & 0 & 0
\end{pmatrix} \end{math}
&
\begin{math} \begin{pmatrix}
0 & -t_{xy} & 0 \\
t_{xy} & 0 & 0 \\
0 & 0 & 0
\end{pmatrix} \end{math}
&
\spacegroup{4/mmm1'}
&
\begin{math} \begin{pmatrix}
\sigma_{xx} & 0 & 0 \\
0 & \sigma_{xx} & 0 \\
0 & 0 & \sigma_{zz}
\end{pmatrix} \end{math}
&
\begin{math} \begin{pmatrix}
0 & \sigma_{xy}^{z} & 0 \\
-\sigma_{xy}^{z} & 0 & 0 \\
0 & 0 & 0
\end{pmatrix} \end{math}
\\[3ex]
\spacegroup{4'/m'm'm}
&
\begin{math} \begin{pmatrix}
t_{xx} & 0 & 0 \\
0 & -t_{xx} & 0 \\
0 & 0 & 0
\end{pmatrix} \end{math}
&
\begin{math} \begin{pmatrix}
t_{xx} & 0 & 0 \\
0 & -t_{xx} & 0 \\
0 & 0 & 0
\end{pmatrix} \end{math}
&
\spacegroup{4/mmm1'}
&
\begin{math} \begin{pmatrix}
\sigma_{xx} & 0 & 0 \\
0 & \sigma_{xx} & 0 \\
0 & 0 & \sigma_{zz}
\end{pmatrix} \end{math}
&
\begin{math} \begin{pmatrix}
0 & \sigma_{xy}^{z} & 0 \\
-\sigma_{xy}^{z} & 0 & 0 \\
0 & 0 & 0
\end{pmatrix} \end{math}
\\[3ex]
\end{tabular}

 \caption{\label{TAB-TORTENc1b} Table~\ref{TAB-TORTENc} continued for tetragonal groups with $O(k) = 4$.}
\end{table*}
%
%
%
\begin{table*}
  \tabulinesep\matrixtablelinesep
  \renewcommand{\arraystretch}{\matrixtablearraystretch}
  \setlength\arraycolsep{\matrixtablearraycolsep}
  \centering
\begin{tabular}{>{\centering}m{5.0cm}cccccc}
\tableheading{magnetic point group} & \tableheading{\TORTEN} & \tableheading{\TORTEN$^\prime$} & \tableheading{magnetic Laue group} & \tableheading{\SIGTEN} & \tableheading{\SIGTEN$^k$} \\[4mm]
\spacegroup{\bar 3'}
&
\begin{math} \begin{pmatrix}
t_{xx} & t_{xy} & 0 \\
-t_{xy} & t_{xx} & 0 \\
0 & 0 & t_{zz}
\end{pmatrix} \end{math}
&
\begin{math} \begin{pmatrix}
t_{xx} & -t_{xy} & 0 \\
t_{xy} & t_{xx} & 0 \\
0 & 0 & t_{zz}
\end{pmatrix} \end{math}
&
\spacegroup{\bar 31'}
&
\begin{math} \begin{pmatrix}
\sigma_{xx} & 0 & 0 \\
0 & \sigma_{xx} & 0 \\
0 & 0 & \sigma_{zz}
\end{pmatrix} \end{math}
&
\begin{math} \begin{pmatrix}
\sigma_{xx}^{z} & \sigma_{xy}^{z} & 0 \\
-\sigma_{xy}^{z} & \sigma_{xx}^{z} & 0 \\
0 & 0 & \sigma_{zz}^{z}
\end{pmatrix} \end{math}
\\[3ex]
\spacegroup{312'}
&
\begin{math} \begin{pmatrix}
t_{xx} & t_{xy} & 0 \\
-t_{xy} & t_{xx} & 0 \\
0 & 0 & t_{zz}
\end{pmatrix} \end{math}
&
\begin{math} \begin{pmatrix}
-t_{xx} & -t_{xy} & 0 \\
t_{xy} & -t_{xx} & 0 \\
0 & 0 & -t_{zz}
\end{pmatrix} \end{math}
&
\spacegroup{\bar 31m'}
&
\begin{math} \begin{pmatrix}
\sigma_{xx} & \sigma_{xy} & 0 \\
-\sigma_{xy} & \sigma_{xx} & 0 \\
0 & 0 & \sigma_{zz}
\end{pmatrix} \end{math}
&
\begin{math} \begin{pmatrix}
\sigma_{xx}^{z} & \sigma_{xy}^{z} & 0 \\
-\sigma_{xy}^{z} & \sigma_{xx}^{z} & 0 \\
0 & 0 & \sigma_{zz}^{z}
\end{pmatrix} \end{math}
\\[3ex]
\spacegroup{31m'}
&
\begin{math} \begin{pmatrix}
t_{xx} & t_{xy} & 0 \\
-t_{xy} & t_{xx} & 0 \\
0 & 0 & t_{zz}
\end{pmatrix} \end{math}
&
\begin{math} \begin{pmatrix}
t_{xx} & t_{xy} & 0 \\
-t_{xy} & t_{xx} & 0 \\
0 & 0 & t_{zz}
\end{pmatrix} \end{math}
&
\spacegroup{\bar 31m'}
&
\begin{math} \begin{pmatrix}
\sigma_{xx} & \sigma_{xy} & 0 \\
-\sigma_{xy} & \sigma_{xx} & 0 \\
0 & 0 & \sigma_{zz}
\end{pmatrix} \end{math}
&
\begin{math} \begin{pmatrix}
\sigma_{xx}^{z} & \sigma_{xy}^{z} & 0 \\
-\sigma_{xy}^{z} & \sigma_{xx}^{z} & 0 \\
0 & 0 & \sigma_{zz}^{z}
\end{pmatrix} \end{math}
\\[3ex]
\spacegroup{\bar 3'1m'}
&
\begin{math} \begin{pmatrix}
t_{xx} & 0 & 0 \\
0 & t_{xx} & 0 \\
0 & 0 & t_{zz}
\end{pmatrix} \end{math}
&
\begin{math} \begin{pmatrix}
t_{xx} & 0 & 0 \\
0 & t_{xx} & 0 \\
0 & 0 & t_{zz}
\end{pmatrix} \end{math}
&
\spacegroup{\bar 31m1'}
&
\begin{math} \begin{pmatrix}
\sigma_{xx} & 0 & 0 \\
0 & \sigma_{xx} & 0 \\
0 & 0 & \sigma_{zz}
\end{pmatrix} \end{math}
&
\begin{math} \begin{pmatrix}
0 & \sigma_{xy}^{z} & 0 \\
-\sigma_{xy}^{z} & 0 & 0 \\
0 & 0 & 0
\end{pmatrix} \end{math}
\\[3ex]
\spacegroup{\bar 3'1m}
&
\begin{math} \begin{pmatrix}
0 & t_{xy} & 0 \\
-t_{xy} & 0 & 0 \\
0 & 0 & 0
\end{pmatrix} \end{math}
&
\begin{math} \begin{pmatrix}
0 & -t_{xy} & 0 \\
t_{xy} & 0 & 0 \\
0 & 0 & 0
\end{pmatrix} \end{math}
&
\spacegroup{\bar 31m1'}
&
\begin{math} \begin{pmatrix}
\sigma_{xx} & 0 & 0 \\
0 & \sigma_{xx} & 0 \\
0 & 0 & \sigma_{zz}
\end{pmatrix} \end{math}
&
\begin{math} \begin{pmatrix}
0 & \sigma_{xy}^{z} & 0 \\
-\sigma_{xy}^{z} & 0 & 0 \\
0 & 0 & 0
\end{pmatrix} \end{math}
\\[3ex]
\end{tabular}

 \caption{\label{TAB-TORTENc2a} Table~\ref{TAB-TORTENc} continued for trigonal groups with $O(k) = 3$.}
\end{table*}
%
%
%
\begin{table*}
  \tabulinesep\matrixtablelinesep
  \renewcommand{\arraystretch}{\matrixtablearraystretch}
  \setlength\arraycolsep{\matrixtablearraycolsep}
  \centering
\begin{tabular}{>{\centering}m{5.0cm}cccccc}
\tableheading{magnetic point group} & \tableheading{\TORTEN} & \tableheading{\TORTEN$^\prime$} & \tableheading{magnetic Laue group} & \tableheading{\SIGTEN} & \tableheading{\SIGTEN$^k$} \\[4mm]
\spacegroup{6'}
&
\begin{math} \begin{pmatrix}
t_{xx} & t_{xy} & 0 \\
-t_{xy} & t_{xx} & 0 \\
0 & 0 & t_{zz}
\end{pmatrix} \end{math}
&
\begin{math} \begin{pmatrix}
-t_{xx} & t_{xy} & 0 \\
-t_{xy} & -t_{xx} & 0 \\
0 & 0 & -t_{zz}
\end{pmatrix} \end{math}
&
\spacegroup{6'/m'}
&
\begin{math} \begin{pmatrix}
\sigma_{xx} & 0 & 0 \\
0 & \sigma_{xx} & 0 \\
0 & 0 & \sigma_{zz}
\end{pmatrix} \end{math}
&
\begin{math} \begin{pmatrix}
\sigma_{xx}^{z} & \sigma_{xy}^{z} & 0 \\
-\sigma_{xy}^{z} & \sigma_{xx}^{z} & 0 \\
0 & 0 & \sigma_{zz}^{z}
\end{pmatrix} \end{math}
\\[3ex]
\spacegroup{\bar 6'}
&
\begin{math} \begin{pmatrix}
t_{xx} & t_{xy} & 0 \\
-t_{xy} & t_{xx} & 0 \\
0 & 0 & t_{zz}
\end{pmatrix} \end{math}
&
\begin{math} \begin{pmatrix}
t_{xx} & -t_{xy} & 0 \\
t_{xy} & t_{xx} & 0 \\
0 & 0 & t_{zz}
\end{pmatrix} \end{math}
&
\spacegroup{6'/m'}
&
\begin{math} \begin{pmatrix}
\sigma_{xx} & 0 & 0 \\
0 & \sigma_{xx} & 0 \\
0 & 0 & \sigma_{zz}
\end{pmatrix} \end{math}
&
\begin{math} \begin{pmatrix}
\sigma_{xx}^{z} & \sigma_{xy}^{z} & 0 \\
-\sigma_{xy}^{z} & \sigma_{xx}^{z} & 0 \\
0 & 0 & \sigma_{zz}^{z}
\end{pmatrix} \end{math}
\\[3ex]
\spacegroup{6/m'}
&
\begin{math} \begin{pmatrix}
t_{xx} & t_{xy} & 0 \\
-t_{xy} & t_{xx} & 0 \\
0 & 0 & t_{zz}
\end{pmatrix} \end{math}
&
\begin{math} \begin{pmatrix}
t_{xx} & -t_{xy} & 0 \\
t_{xy} & t_{xx} & 0 \\
0 & 0 & t_{zz}
\end{pmatrix} \end{math}
&
\spacegroup{6/m1'}
&
\begin{math} \begin{pmatrix}
\sigma_{xx} & 0 & 0 \\
0 & \sigma_{xx} & 0 \\
0 & 0 & \sigma_{zz}
\end{pmatrix} \end{math}
&
\begin{math} \begin{pmatrix}
\sigma_{xx}^{z} & \sigma_{xy}^{z} & 0 \\
-\sigma_{xy}^{z} & \sigma_{xx}^{z} & 0 \\
0 & 0 & \sigma_{zz}^{z}
\end{pmatrix} \end{math}
\\[3ex]
\spacegroup{6'22'}
&
\begin{math} \begin{pmatrix}
t_{xx} & 0 & 0 \\
0 & t_{xx} & 0 \\
0 & 0 & t_{zz}
\end{pmatrix} \end{math}
&
\begin{math} \begin{pmatrix}
-t_{xx} & 0 & 0 \\
0 & -t_{xx} & 0 \\
0 & 0 & -t_{zz}
\end{pmatrix} \end{math}
&
\spacegroup{6'/m'mm'}
&
\begin{math} \begin{pmatrix}
\sigma_{xx} & 0 & 0 \\
0 & \sigma_{xx} & 0 \\
0 & 0 & \sigma_{zz}
\end{pmatrix} \end{math}
&
\begin{math} \begin{pmatrix}
0 & \sigma_{xy}^{z} & 0 \\
-\sigma_{xy}^{z} & 0 & 0 \\
0 & 0 & 0
\end{pmatrix} \end{math}
\\[3ex]
\spacegroup{62'2'}
&
\begin{math} \begin{pmatrix}
t_{xx} & t_{xy} & 0 \\
-t_{xy} & t_{xx} & 0 \\
0 & 0 & t_{zz}
\end{pmatrix} \end{math}
&
\begin{math} \begin{pmatrix}
-t_{xx} & -t_{xy} & 0 \\
t_{xy} & -t_{xx} & 0 \\
0 & 0 & -t_{zz}
\end{pmatrix} \end{math}
&
\spacegroup{6/mm'm'}
&
\begin{math} \begin{pmatrix}
\sigma_{xx} & \sigma_{xy} & 0 \\
-\sigma_{xy} & \sigma_{xx} & 0 \\
0 & 0 & \sigma_{zz}
\end{pmatrix} \end{math}
&
\begin{math} \begin{pmatrix}
\sigma_{xx}^{z} & \sigma_{xy}^{z} & 0 \\
-\sigma_{xy}^{z} & \sigma_{xx}^{z} & 0 \\
0 & 0 & \sigma_{zz}^{z}
\end{pmatrix} \end{math}
\\[3ex]
\spacegroup{6'mm'}
&
\begin{math} \begin{pmatrix}
0 & t_{xy} & 0 \\
-t_{xy} & 0 & 0 \\
0 & 0 & 0
\end{pmatrix} \end{math}
&
\begin{math} \begin{pmatrix}
0 & t_{xy} & 0 \\
-t_{xy} & 0 & 0 \\
0 & 0 & 0
\end{pmatrix} \end{math}
&
\spacegroup{6'/m'mm'}
&
\begin{math} \begin{pmatrix}
\sigma_{xx} & 0 & 0 \\
0 & \sigma_{xx} & 0 \\
0 & 0 & \sigma_{zz}
\end{pmatrix} \end{math}
&
\begin{math} \begin{pmatrix}
0 & \sigma_{xy}^{z} & 0 \\
-\sigma_{xy}^{z} & 0 & 0 \\
0 & 0 & 0
\end{pmatrix} \end{math}
\\[3ex]
\spacegroup{6m'm'}
&
\begin{math} \begin{pmatrix}
t_{xx} & t_{xy} & 0 \\
-t_{xy} & t_{xx} & 0 \\
0 & 0 & t_{zz}
\end{pmatrix} \end{math}
&
\begin{math} \begin{pmatrix}
t_{xx} & t_{xy} & 0 \\
-t_{xy} & t_{xx} & 0 \\
0 & 0 & t_{zz}
\end{pmatrix} \end{math}
&
\spacegroup{6/mm'm'}
&
\begin{math} \begin{pmatrix}
\sigma_{xx} & \sigma_{xy} & 0 \\
-\sigma_{xy} & \sigma_{xx} & 0 \\
0 & 0 & \sigma_{zz}
\end{pmatrix} \end{math}
&
\begin{math} \begin{pmatrix}
\sigma_{xx}^{z} & \sigma_{xy}^{z} & 0 \\
-\sigma_{xy}^{z} & \sigma_{xx}^{z} & 0 \\
0 & 0 & \sigma_{zz}^{z}
\end{pmatrix} \end{math}
\\[3ex]
\spacegroup{\bar 6'2m'}
&
\begin{math} \begin{pmatrix}
t_{xx} & 0 & 0 \\
0 & t_{xx} & 0 \\
0 & 0 & t_{zz}
\end{pmatrix} \end{math}
&
\begin{math} \begin{pmatrix}
t_{xx} & 0 & 0 \\
0 & t_{xx} & 0 \\
0 & 0 & t_{zz}
\end{pmatrix} \end{math}
&
\spacegroup{6'/m'mm'}
&
\begin{math} \begin{pmatrix}
\sigma_{xx} & 0 & 0 \\
0 & \sigma_{xx} & 0 \\
0 & 0 & \sigma_{zz}
\end{pmatrix} \end{math}
&
\begin{math} \begin{pmatrix}
0 & \sigma_{xy}^{z} & 0 \\
-\sigma_{xy}^{z} & 0 & 0 \\
0 & 0 & 0
\end{pmatrix} \end{math}
\\[3ex]
\spacegroup{\bar 6'm2'}
&
\begin{math} \begin{pmatrix}
0 & t_{xy} & 0 \\
-t_{xy} & 0 & 0 \\
0 & 0 & 0
\end{pmatrix} \end{math}
&
\begin{math} \begin{pmatrix}
0 & -t_{xy} & 0 \\
t_{xy} & 0 & 0 \\
0 & 0 & 0
\end{pmatrix} \end{math}
&
\spacegroup{6'/m'mm'}
&
\begin{math} \begin{pmatrix}
\sigma_{xx} & 0 & 0 \\
0 & \sigma_{xx} & 0 \\
0 & 0 & \sigma_{zz}
\end{pmatrix} \end{math}
&
\begin{math} \begin{pmatrix}
0 & \sigma_{xy}^{z} & 0 \\
-\sigma_{xy}^{z} & 0 & 0 \\
0 & 0 & 0
\end{pmatrix} \end{math}
\\[3ex]
\spacegroup{6/m'm'm'}
&
\begin{math} \begin{pmatrix}
t_{xx} & 0 & 0 \\
0 & t_{xx} & 0 \\
0 & 0 & t_{zz}
\end{pmatrix} \end{math}
&
\begin{math} \begin{pmatrix}
t_{xx} & 0 & 0 \\
0 & t_{xx} & 0 \\
0 & 0 & t_{zz}
\end{pmatrix} \end{math}
&
\spacegroup{6/mmm1'}
&
\begin{math} \begin{pmatrix}
\sigma_{xx} & 0 & 0 \\
0 & \sigma_{xx} & 0 \\
0 & 0 & \sigma_{zz}
\end{pmatrix} \end{math}
&
\begin{math} \begin{pmatrix}
0 & \sigma_{xy}^{z} & 0 \\
-\sigma_{xy}^{z} & 0 & 0 \\
0 & 0 & 0
\end{pmatrix} \end{math}
\\[3ex]
\spacegroup{6/m'mm}
&
\begin{math} \begin{pmatrix}
0 & t_{xy} & 0 \\
-t_{xy} & 0 & 0 \\
0 & 0 & 0
\end{pmatrix} \end{math}
&
\begin{math} \begin{pmatrix}
0 & -t_{xy} & 0 \\
t_{xy} & 0 & 0 \\
0 & 0 & 0
\end{pmatrix} \end{math}
&
\spacegroup{6/mmm1'}
&
\begin{math} \begin{pmatrix}
\sigma_{xx} & 0 & 0 \\
0 & \sigma_{xx} & 0 \\
0 & 0 & \sigma_{zz}
\end{pmatrix} \end{math}
&
\begin{math} \begin{pmatrix}
0 & \sigma_{xy}^{z} & 0 \\
-\sigma_{xy}^{z} & 0 & 0 \\
0 & 0 & 0
\end{pmatrix} \end{math}
\\[3ex]
\end{tabular}

 \caption{\label{TAB-TORTENc2b} Table~\ref{TAB-TORTENc} continued for hexagonal groups with $O(k) = 6$.}
\end{table*}
%
%
%
\begin{table*}
  \tabulinesep\matrixtablelinesep
  \renewcommand{\arraystretch}{\matrixtablearraystretch}
  \setlength\arraycolsep{\matrixtablearraycolsep}
  \centering
\begin{tabular}{ccccccc}
\tableheading{magnetic point group} & \tableheading{\TORTEN} & \tableheading{\TORTEN$^\prime$} & \tableheading{magnetic Laue group} & \tableheading{\SIGTEN} & \tableheading{\SIGTEN$^k$} \\[4mm]
\spacegroup{m'\bar 3'}
&
\begin{math} \begin{pmatrix}
t_{xx} & 0 & 0 \\
0 & t_{xx} & 0 \\
0 & 0 & t_{xx}
\end{pmatrix} \end{math}
&
\begin{math} \begin{pmatrix}
t_{xx} & 0 & 0 \\
0 & t_{xx} & 0 \\
0 & 0 & t_{xx}
\end{pmatrix} \end{math}
&
\spacegroup{m\bar 31'}
&
\begin{math} \begin{pmatrix}
\sigma_{xx} & 0 & 0 \\
0 & \sigma_{xx} & 0 \\
0 & 0 & \sigma_{xx}
\end{pmatrix} \end{math}
&
\begin{math} \begin{pmatrix}
0 & \sigma_{xy}^{z} & 0 \\
\sigma_{xz}^{y} & 0 & 0 \\
0 & 0 & 0
\end{pmatrix} \end{math}
\\[3ex]
\spacegroup{4'32'}
&
\begin{math} \begin{pmatrix}
t_{xx} & 0 & 0 \\
0 & t_{xx} & 0 \\
0 & 0 & t_{xx}
\end{pmatrix} \end{math}
&
\begin{math} \begin{pmatrix}
-t_{xx} & 0 & 0 \\
0 & -t_{xx} & 0 \\
0 & 0 & -t_{xx}
\end{pmatrix} \end{math}
&
\spacegroup{m\bar 3m'}
&
\begin{math} \begin{pmatrix}
\sigma_{xx} & 0 & 0 \\
0 & \sigma_{xx} & 0 \\
0 & 0 & \sigma_{xx}
\end{pmatrix} \end{math}
&
\begin{math} \begin{pmatrix}
0 & \sigma_{xy}^{z} & 0 \\
\sigma_{xz}^{y} & 0 & 0 \\
0 & 0 & 0
\end{pmatrix} \end{math}
\\[3ex]
\spacegroup{\bar 4'3m'}
&
\begin{math} \begin{pmatrix}
t_{xx} & 0 & 0 \\
0 & t_{xx} & 0 \\
0 & 0 & t_{xx}
\end{pmatrix} \end{math}
&
\begin{math} \begin{pmatrix}
t_{xx} & 0 & 0 \\
0 & t_{xx} & 0 \\
0 & 0 & t_{xx}
\end{pmatrix} \end{math}
&
\spacegroup{m\bar 3m'}
&
\begin{math} \begin{pmatrix}
\sigma_{xx} & 0 & 0 \\
0 & \sigma_{xx} & 0 \\
0 & 0 & \sigma_{xx}
\end{pmatrix} \end{math}
&
\begin{math} \begin{pmatrix}
0 & \sigma_{xy}^{z} & 0 \\
\sigma_{xz}^{y} & 0 & 0 \\
0 & 0 & 0
\end{pmatrix} \end{math}
\\[3ex]
\spacegroup{m'\bar 3'm'}
&
\begin{math} \begin{pmatrix}
t_{xx} & 0 & 0 \\
0 & t_{xx} & 0 \\
0 & 0 & t_{xx}
\end{pmatrix} \end{math}
&
\begin{math} \begin{pmatrix}
t_{xx} & 0 & 0 \\
0 & t_{xx} & 0 \\
0 & 0 & t_{xx}
\end{pmatrix} \end{math}
&
\spacegroup{m\bar 3m1'}
&
\begin{math} \begin{pmatrix}
\sigma_{xx} & 0 & 0 \\
0 & \sigma_{xx} & 0 \\
0 & 0 & \sigma_{xx}
\end{pmatrix} \end{math}
&
\begin{math} \begin{pmatrix}
0 & \sigma_{xy}^{z} & 0 \\
-\sigma_{xy}^{z} & 0 & 0 \\
0 & 0 & 0
\end{pmatrix} \end{math}
\\[3ex]
\end{tabular}

 \caption{\label{TAB-TORTENc2c} Table~\ref{TAB-TORTENc} continued for cubic groups.}
\end{table*}
%

\section{Results \label{sec:R}}

\begin{figure}
 \begin{center}                                                                                                    
 \includegraphics[angle=0,width=0.8\linewidth,clip]{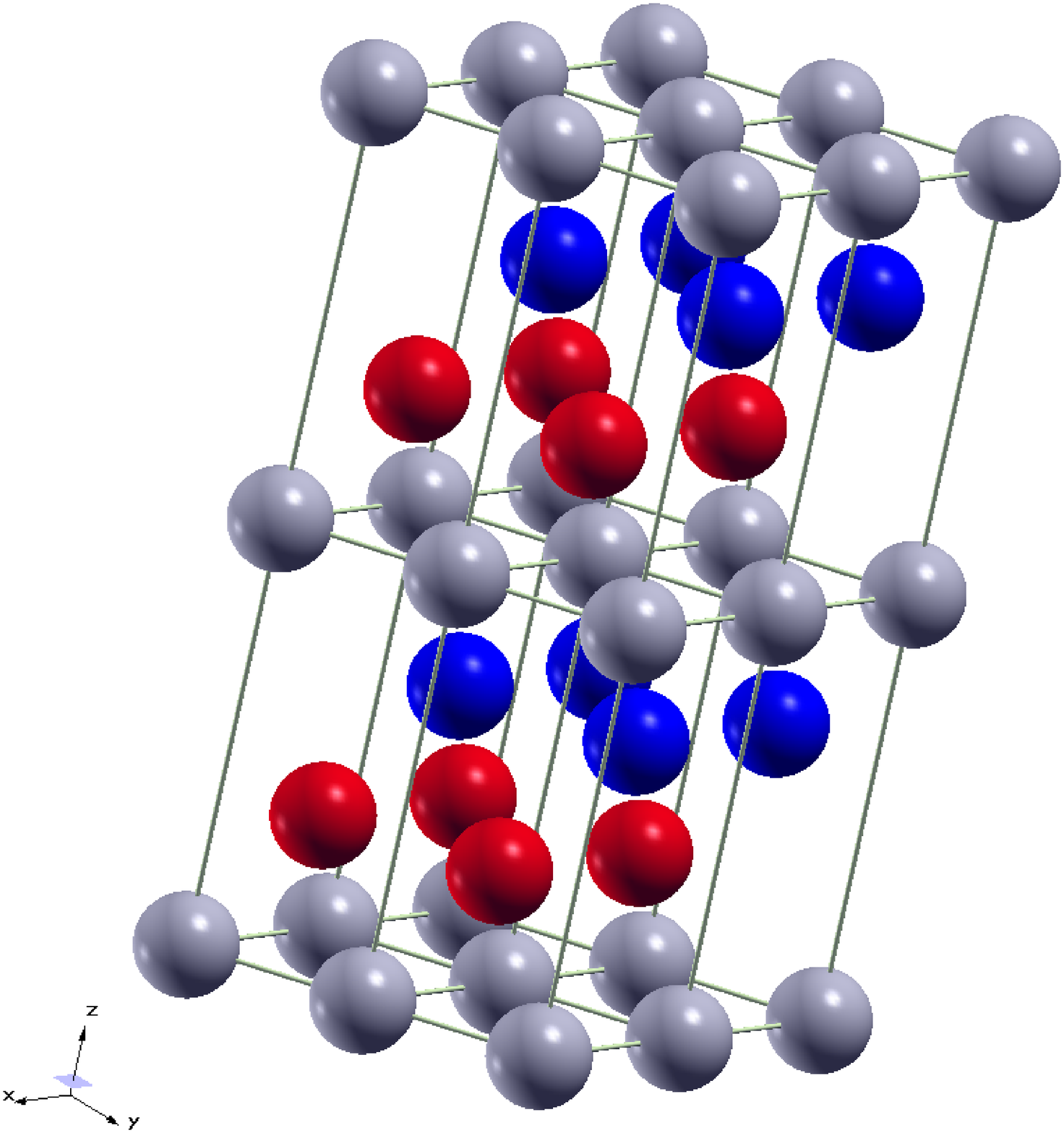}
   \caption{\label{FIG:PtFeCoCu} (Color online) Structure of the investigated
multilayer system \ptfecocu consting of a stacking of   fcc (111)
planes  along the z axis. Cu atoms are colored in blue, Fe$_x$Co$_{1-x}$ sites
in red and Pt atoms  are represented in  light grey.}                                   
\end{center}                                                                                                       
\end{figure}

To investigate the impact of chemical disorder and the \sw{ability to tailor}
the torkance via the alloy composition the multilayer system \ptfecocu has been
investigated over the full range of concentration $x$.  Fig.~\ref{FIG:PtFeCoCu}
shows the hexagonal structure  of the model system for which a stacking of  fcc
(111)-like atomic planes along the z axis has been assumed.
%
%

\begin{figure}
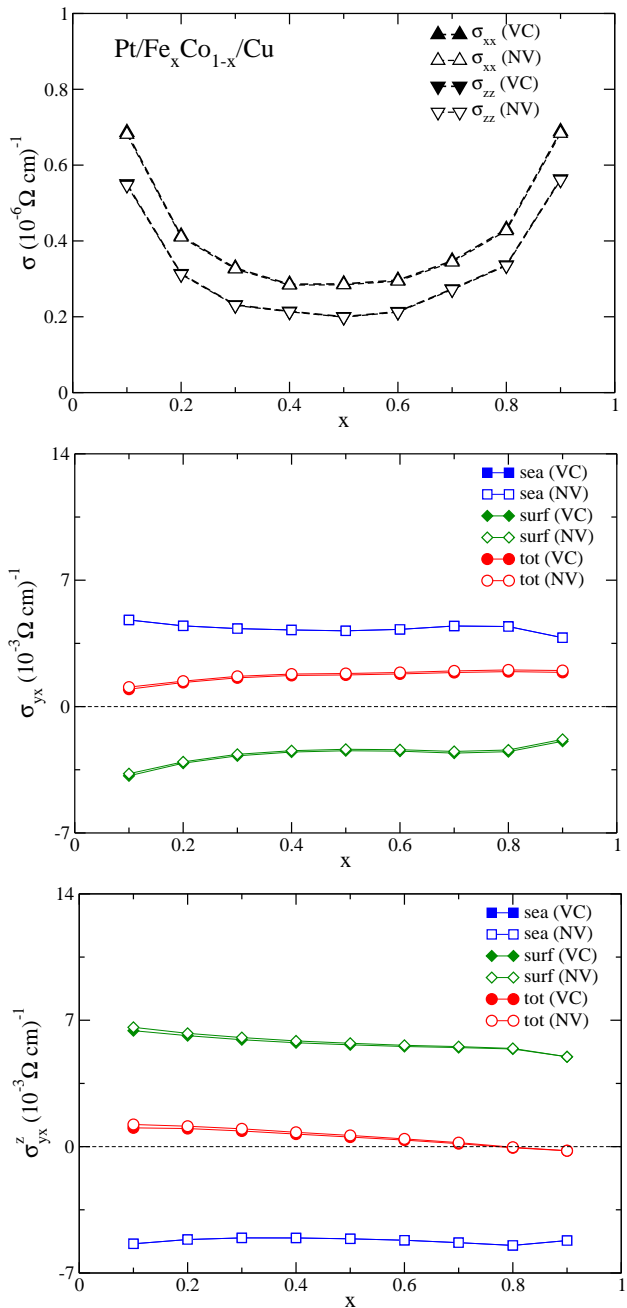

 \begin{center}
                \includegraphics[angle=0,height=0.65\linewidth,clip]{Pt_FeCo_Cu_sig_xx_zz.eps} \\
\vspace{0.2cm}
                \includegraphics[angle=0,height=0.65\linewidth,clip]{Pt_FeCo_Cu_sig_yx_AHE.eps} \\ 
\vspace{0.2cm}
\hspace{0.01cm} \includegraphics[angle=0,height=0.65\linewidth,clip]{Pt_FeCo_Cu_sig_yx_SHE.eps}
   \caption{\label{FIG:SIGMA} (Color online) Top: The longitudinal components
$\sigma_{\rm xx}=\sigma_{\rm yy}$ and $\sigma_{\rm zz}$ of the conductivity
tensor \SIGTEN\ of \ptfecocu as a function of the concentration
$x$.  Middle:  The corresponding \sw{anomalous} Hall conductivity $\sigma_{\rm
xy}=-\sigma_{\rm yx}$.  Bottom: The spin Hall conductivity $\sigma_{\rm
xy}^{\rm z}=-\sigma_{\rm yx}^{\rm z}$.  \sw{Open symbols represent
calculations without vertex corrections (NV) and filled symbols those including
vertex corrections (VC).  The blue squares correspond to the Fermi sea
contribution (sea)\if{} \tr{(sum of 0-term and 1-term, NOT INTRODUCED!?)}\fi, the
green diamonds represent contributions from the Fermi surface (surf) \if{}\tr{(sum of
0-term and 1-term, NOT INTRODUCED!?)}\fi and red circles give the total result (tot).}}
\end{center}
\vspace*{-0.75cm}
\end{figure}

To examine the connection of the  torkance   with other related response
quantities we calculated the \sw{electrical and spin conductivity tensors} in
addition. Replacing the torque operator $\hat{T}_\mu$ in
Eq.~\eqref{eq:torkance-Bastin1} \sw{ by the operator $\hat{j}_\mu$} one gets,
apart from some constants, the corresponding expressions for the electrical
conductivity tensor \SIGTEN. From this one can see immediately that the
longitudinal conductivities $\sigma_{ii}$ are connected only with the first
Fermi surface term in Eq.~\eqref{eq:torkance-Bastin1}, accordingly they are
determined for $T=   0$~K by the electronic structure at the Fermi energy
$E_{\rm F}$  while the second Fermi sea term vanishes.  Due to the magnetic
Laue group (\spacegroup{\bar3m'}) of the investigated system the conductivity
tensor \SIGTEN\ has only the non-vanishing elements \sw{$\sigma_{\rm
xx}=\sigma_{\rm yy}\ne \sigma_{\rm zz}$ and $\sigma_{\rm xy}=- \sigma_{\rm
yx}$,\cite{SKWE15} i.e., the well-known shape of ferromagnetic systems with a
principal axis $k$ of order $O(k) \ge 3$ and neither additional purely spatial
rotation axes perpendicular to it, nor vertical mirror planes.
} 
The corresponding results for \ptfecocu are shown in Fig.~\ref{FIG:SIGMA} as a
function of the concentration $x$. 
%
%
As to be expected for $T=0$~K one finds a divergent behavior for the
longitudinal conductivities $\sigma_{\rm xx}$ and $\sigma_{\rm zz}$ in the
dilute regime, i.e., when $x$ goes to 0 or 1, respectively. In both cases the
variation with $x$ is rather symmetric around the composition $x=0.5$ as the
two alloying components, Fe and Co, respectively, do not differ too much
concerning their electronic properties \sw{\tr{in this fcc (111)-like
structure.}} Apart from this general behavior one notes that one has
$\sigma_{\rm xx} > \sigma_{\rm zz}$ for all concentrations.  This is due to the
simple fact that for $\sigma_{\rm xx}$ one has electronic transport parallel to
the atomic layers while $\sigma_{\rm zz}$ implies transport perpendicular to
the layers, the finite conductivity is not only because of the chemical
disorder in the Fe-Co layers but in addition due to a strong geometrical
confinement \sw{and corresponding interface scattering}. Fig.~\ref{FIG:SIGMA}
(top) shows also the conductivity $\sigma_{\rm xx}$ and $\sigma_{\rm zz}$
calculated without the vertex corrections. As one can see, this restriction
hardly changes the numerical results.  This finding is very typical for
transition metal systems with a high density of states at the Fermi energy
implying a short mean free path length.\cite{BEV97} In contrast to the
longitudinal conductivity $\sigma_{ii}$ the transverse conductivity
$\sigma_{\rm xy}$ has contributions from the Fermi surface as well as Fermi sea
terms (see Eq.~\eqref{eq:torkance-Bastin1}) when the  Kubo-Bastin formula is
used (see comment below).  Corresponding results for \ptfecocu are shown in the
middle panel of Fig.~\ref{FIG:SIGMA}. As one notes, the Fermi surface and sea
contributions are comparable in magnitude but have opposite sign leading to a
partial cancellation.  Obviously, both contributions vary rather smoothly with
concentration and show for the considered concentration range ($0.1 \le x \le
0.9$) in contrast for example to the binary alloys Fe$_x$Pd$_{1-x}$ and
Ni$_x$Pd$_{1-x}$ \cite{LKE10b}  practically no divergent behavior in the dilute
limit ($x\rightarrow  0$ or $x\rightarrow  1$). As discussed before
\cite{LKE10b} a divergent behavior of $\sigma_{\rm xy}$ can be ascribed to a
strong skew scattering contribution that scales with the longitudinal
conductivity $\sigma_{\rm xx}$.\cite{NSO+10}  On the other hand, this extrinsic
source for the transverse transport is accounted for by the contribution to
$\sigma_{\rm xy}$ that is connected with the vertex corrections.\cite{LKE10b}
Inspecting Fig.~\ref{FIG:SIGMA}  (middle) that shows results for the Fermi
surface contribution to $\sigma_{\rm xy}$ obtained with and without the vertex
corrections, one finds that these give rise only to minor corrections
throughout the considered concentration regime. With the skew scattering
mechanism being negligible and the intrinsic contribution dominating the system
is obviously in the so-called dirty regime.\cite{NSO+10,BRS14} Considering the
Fermi sea contribution to $\sigma_{\rm xy}$ (Fig.~\ref{FIG:SIGMA} middle) one
finds no impact of the vertex corrections at all.  This is in full line with
the findings of Turek \ea\cite{TKD14} who could show (at least within the
TB-LMTO-CPA formalism) that this property has to be fulfilled for formal
reasons. As a consequence, this implies that the skew scattering mechanism is,
as to be expected, connected only to the Fermi surface contribution to
$\sigma_{\rm xy}$. In fact, this is a seemingly trivial precondition to get the
full skew scattering contribution to $\sigma_{\rm xy}$ when performing
Boltzmann type of calculations for the dilute regime that are restricted to the
Fermi energy $E_{\rm F}$ .\cite{ZCK+14}  In fact, this is to be expected
because for the electrical conductivity tensor it is possible for the case
$T=0$~K to go from the Kubo-Bastin to the Kubo-St\v{r}eda  equation that has
only contributions from the Fermi surface,\cite{Str82,CB01a} i.e., the Fermi
sea term can be eliminated exactly.

Considering the spin conductivity tensor the non-vanishing tensor elements
$\sigma_{ij}^{k}$  can again be found from symmetry considerations.
\cite{SKWE15}  Restricting here to the z component of the spin polarization
one has the non-vanishing elements \sw{$\sigma_{\rm xx}^{\rm z} = \sigma_{\rm
yy}^{\rm z} \neq \sigma_{\rm zz}^{\rm z}$ and $\sigma_{\rm xy}^{\rm z}
=-\sigma_{\rm yx}^{\rm z}$, i.e., \SIGTEN$^z$ has the same shape as \SIGTEN.}
Comparing the corresponding numerical results for the transverse spin
conductivity shown in the lower panel of Fig.~\ref{FIG:SIGMA} with their
counterparts connected with the transverse conductivity $\sigma_{\rm xy}$ one
finds that a very similar behavior in the investigated concentration regime: i)
the Fermi sea and surface contributions are comparable in magnitude but have
different sign leading to a pronounced cancellation, ii) the individual terms
vary very weakly with concentration without showing any divergent behavior,
iii) the Fermi surface contribution shows a very weak impact of the vertex
contributions while, iv) the Fermi sea contribution is not affected at all by
the vertex corrections.  The findings iii) and iv) again imply that the
extrinsic contributions and with this the skew scattering contribution are very
small.  Finding iv) that so far has been demonstrated only  numerically is now
(in contact to the case of the electrical conductivity) by no means trivial.
While the use of a Kubo-St\v{r}eda-like equation for $\sigma_{\rm xy}^{\rm z}$
turned out to be very successful when applied to metallic alloys \cite{LGK+11}
it has to be seen as approximate.\cite{KCE15}  For that reason the finding
that there are no vertex corrections to the Fermi sea part but only for the
Fermi surface part is now an important precondition for getting all skew
scattering contributions to $\sigma_{\rm xy}^{\rm z}$ by performing
calculations based on the Boltzmann equation.\cite{GFZM10,HFM+13,ZCK+14}

\medskip

\begin{figure}
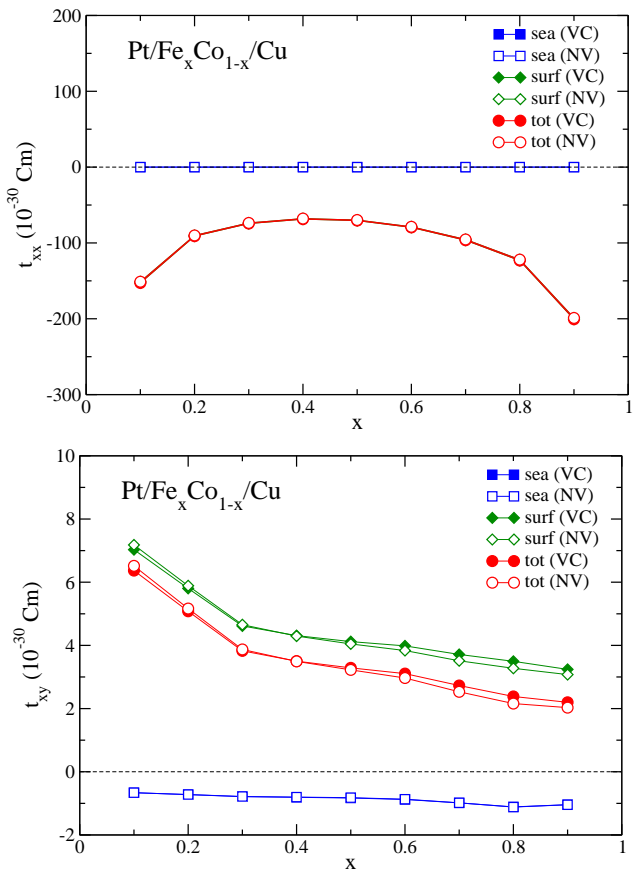

 \begin{center}
                \includegraphics[angle=0,height=0.65\linewidth,clip]{Pt_FeCo_Cu_SOT_xx_SI.eps} \\
\vspace{0.2cm}
                \includegraphics[angle=0,height=0.65\linewidth,clip]{Pt_FeCo_Cu_SOT_xy_SI.eps} \\ 
\vspace{0.2cm}
   \caption{\label{FIG:TORQUE} (Color online) Top: The longitudinal component
$t_{\rm xx}=t_{\rm yy}$ of the SOT depending on the concentration.  Bottom: The
transverse component $t_{\rm xy}=-t_{\rm yx}$ of the SOT depending on the
concentration. \sw{Use of symbols and colors as in Fig.~\ref{FIG:SIGMA}.}}
\end{center}
\end{figure}

For the magnetization along the  z axis \ptfecocu has the magnetic point group
\spacegroup{3m'}\footnote{\sw{More precisely \spacegroup{31m'} or
\spacegroup{3m'1}, depending on the axis convention for the corresponding space
group. Results for the former are given here, the tensors for the other can be
obtained by a rotation of the coordinate system by $\pi/2$ around the principal
axis. See Ref.~\onlinecite{SKWE15} for details}} leading to an anti-symmetric
torkance tensor with non-vanishing elements $ t_{\rm xx} = t_{\rm yy} \ne
t_{\rm zz} $ and $ t_{\rm xy}=  - t_{\rm yx} $ (see row three of
Tab.~\ref{TAB-TORTENc2a}).
Actually, because of the restrictions imposed by the form of the torque
operator given in Eq.~\eqref{matrix-element} the element $t_{\rm zz}$ that
would represent a change in the magnitude of the magnetic moment along the
z direction does not show up in the calculations.  In fact, this impact of an
external electric field \sw{can be considered as a manifestation of the}
Edelstein effect\cite{AL89,Ede90} and can be \sw{described} by a response
quantity formulated appropriately.\footnote{\sw{The close connection between
spin-orbit torques and the Edelstein effect has already been mentioned
earlier\cite{FBM15,LGZ+15,SOC+15} and will be discussed in detail
elsewhere\cite{WCS+16a}}} The top panel of Fig.~\ref{FIG:TORQUE} gives the
numerical results for the diagonal  torkance element $t_{\rm xx}$.
%
%
%
As one can see, it has many properties in common with the longitudinal
conductivity $\sigma_{\rm xx}$: i) there is no Fermi sea contribution, ii) it
shows a divergent behavior in the dilute limit $x\rightarrow 0$ or
$x\rightarrow 1$, respectively.  In contrast to $\sigma_{\rm xx}$, however, we
find no impact of the vertex corrections at all.  This implies that there are
no contributions due to skew scattering and accordingly there is only an
intrinsic contribution to $t_{\rm xx}$.  As a consequence, this  torkance
tensor element will not be accessible by calculations based on the Boltzmann
formalism.  Considering $t_{\rm xy}$ one finds from Figs.~\ref{FIG:SIGMA}  and
\ref{FIG:TORQUE} that this tensor element behaves much like $\sigma_{\rm xy}$
and  $\sigma_{\rm xy}^{\rm z}$: i) the Fermi sea and surface contributions are
comparable in magnitude but have different sign leading to a partial
cancellation, ii)  both parts are weakly concentration dependent with a more
pronounced variation for the Fermi surface term on the Co-rich side, iii) the
Fermi surface contribution shows a very weak impact of the vertex
contributions, while iv) the Fermi sea contribution is not affected at all by
the vertex contributions.  Again, from iii) and iv) one may conclude that the
extrinsic contributions due to the skew scattering are very small.  In contrast
to $t_{\rm xx}$ calculations based on the Boltzmann formalism should be able to
account for this contribution to $t_{\rm xy}$. \sw{The comparable concentration
dependence of the spin Hall conductivity $\sigma_{xy}^z$ and the even torkance
$t_{xy}$ seems to support previous suggestions that they are intimately
connected.\footnote{\sw{See, e.g., Eq.~(76) of Ref.~\onlinecite{FBM15} for an
explicit relation.}}}

\medskip

The first \emph{ab-initio} investigations on the spin-orbit torque by Freimuth
\ea\cite{FBM14,FBM14a} were dealing among others with Co/Pt(111) having the
same symmetry as the system \ptfecocu considered here.  As shown by these
authors, the mirror planes perpendicular to the atomic layers implies the
$t_{\rm xx}$ and $t_{\rm xy}$ to be \sw{odd and even}, respectively, under
reversal of the magnetization direction, i.e., one has \sw{$t_{\rm xx}(\vec{m})
=  -t_{\rm xx}(-\vec{m})$ and $t_{\rm xy}(\vec{m}) = t_{\rm xy}(-\vec{m})$}.
Our numerical results are fully in line with this basic symmetry restriction.
Freimuth \ea\ also used the Kubo-Bastin formalism, however, with the Green
function represented in terms of Bloch functions and energy eigen values.  This
restricted the investigation to the very dilute limit with the impact of
chemical or structural disorder represented by a broadening parameter $\Gamma$.
Calculating the diagonal torkance element $t_{\rm xx}$ as a function of
$\Gamma$ leads in the limit $\Gamma\rightarrow 0$ to a divergent behavior.
This is obviously in full accordance with the results shown in
Fig.~\ref{FIG:TORQUE} (top) that also show a divergence for the concentration
$x \rightarrow  0$ or $x \rightarrow 1$, implying that the major impact of
disorder on the diagonal torkance   is independent of whether it is accounted
for within the framework of the CPA or the Gaussian disorder
model.\cite{FBM14a}  The same applies also to the off-diagonal element $t_{\rm
xy}$.  While $t_{\rm xy}$ given in Fig.\ \ref{FIG:TORQUE} shows only a weak
variation with concentration in the considered composition regime, $t_{\rm xy}$
of Co/Pt(111) as calculated by Freimuth \ea\cite{FBM14a} as a function of the
broadening parameter takes a constant and finite value in the limit $\Gamma
\rightarrow  0$, the intrinsic contribution to the torkance.  Concerning the
decomposition of the torkance into Fermi sea and Fermi surface contributions,
the results in Fig.~\ref{FIG:TORQUE} are again in qualitative agreement with
the findings of Freimuth \ea\cite{FBM14,FBM14a}: The odd torkance element
$t_{xx}$ (top) has no Fermi sea contribution whereas to the even $t_{xy}$
(bottom) both, Fermi sea and Fermi surface, contribute significantly.
\sw{Finally, as suggested before -- amongst others by the aforementioned
authors -- the similar composition dependence of $t_{xy}$ and the spin Hall
conductivity $\sigma_{xy}^z$ seems to support at least in part the notion
``spin~Hall''-torque.}

\medskip

\section{Conclusions \label{sec:C}}

In summary, based on Kubo's linear response formalism, the symmetry and
magnitude of spin-orbit torques in metals and alloys can be investigated using
group-theoretical considerations for the former and an implementation of the
Kubo-Bastin formula for the torkance in a multiple-scattering framework for the
latter. The resulting tensor shapes for direct and inverse torkance for all
magnetic point groups allowing a finite magnetization have been presented.
\tr{The former have been independently confirmed for a number of systems by
numerical calculations.} By investigating the concentration dependence of two
symmetrically distinct tensor elements in an fcc (111) trilayer system, contact
and extensions could be made to previous work concerning the various
contributions to the SOT and possible underlying mechanisms. While the odd
torkance was found to bear a striking resemblance to the electrical
conductivity concerning its dependence on the alloy composition in the
ferromagnetic layer, the even component could be demonstrated to behave more
like the transverse transport properties anomalous and spin Hall conductivity.
The key advantage of the CPA alloy theory over simpler models of disorder is
the possibility to calculate material-specific parameters very efficiently,
opening the way to a computational materials design approach to direct and
inverse spin-orbit torques. As has been shown, the electronic contribution to
the corresponding thermally-induced phenomena, direct and inverse thermal
spin-orbit torques, can in principle be calculated from the torkance employing
a Mott-like expression. Future work will focus on the close connection between
direct and inverse SOT to direct and inverse Edelstein effect.


\begin{acknowledgments}
The authors would like to thank the {\it Deutsche Forschungsgemeinschaft}
(German Science Foundation, DFG) for financial support via the programmes
SPP~1538 and SFB~689.\\
\end{acknowledgments}


%


\begin{thebibliography}{79}
\bibitem [{FBM()}]{FBM14remark}%
  \BibitemOpen
  \href@noop {} {}\bibinfo {howpublished} {The expression for the torkance
  tensor given here differs slighty from that given by Freimuth \textit{et al.}
  \cite{FBM14,FBM14a} but is in full accordance with the corresponding
  expression for the conductivity tensor as given for example by Cr\'epieux and
  Bruno \cite{CB01a}}\BibitemShut {NoStop}%
\end{thebibliography}
\end{document}